\documentclass[11pt]{article}
\baselineskip
\usepackage{amssymb}
\usepackage{graphicx}
\usepackage{latexsym}
\usepackage{cite}
\usepackage{collref}
\usepackage{epsfig}
\usepackage{amsmath}
\usepackage{bbm}
\usepackage{color}
\usepackage{hyperref}
\newcommand{\Z}{\mathbb{Z}}

\newcommand{\SU}{\mathrm{SU}}

\newcommand{\dd}{{\rm{d}}}

\newcommand{\tr}{{\rm Tr}}
\newcommand{\real}{{\rm Re\,}}

\newcommand{\ide}{\mathbbm{1}}
\newcommand{\qhat}{\hat{q}}
\newcommand{\qhatsoft}{\hat{q}_{\mbox{\tiny{soft}}}}

\newcommand{\gE}{g_{\mbox{\tiny{E}}}}
\newcommand{\gsqE}{g^2_{\mbox{\tiny{E}}}}
\newcommand{\gsqM}{g^2_{\mbox{\tiny{M}}}}
\newcommand{\mE}{m_{\mbox{\tiny{E}}}}
\newcommand{\mD}{m_{\mbox{\tiny{D}}}}
\newcommand{\Cfund}{\mathcal{C}_{\mbox{\tiny{f}}}}
\newcommand{\Cadj}{\mathcal{C}_{\mbox{\tiny{a}}}}
\newcommand{\tfund}{t_{\mbox{\tiny{f}}}}
\newcommand{\qmax}{q_{\mbox{\tiny{max}}}}
\newcommand{\LQCD}{\mathcal{L}_{\tiny\mbox{QCD}}}

\newcommand{\LambdaMSbar}{\Lambda_{\overline{\tiny\mbox{MS}}}}
\newcommand{\nb}{n_{\mbox{\tiny{B}}}}

\newcommand{\eq}{\begin{equation}}
\newcommand{\en}{\end{equation}}
\newcommand{\eqar}{\begin{eqnarray}}
\newcommand{\enar}{\end{eqnarray}}

\begin{document}

\begin{titlepage}
\vskip0.5cm 
\begin{flushright} 
HIP-2013-12/TH\\IFT-UAM/CSIC-13-098
\end{flushright} 
\vskip0.5cm 
\begin{center}
{\Large\bf A lattice study of the jet quenching parameter}
\end{center}
\vskip1.3cm
\centerline{Marco~Panero$^{a,b}$, Kari~Rummukainen$^{b}$ and Andreas~Sch\"afer$^{c}$}
\vskip1.5cm
\centerline{\sl $^a$ Instituto de F\'{\i}sica T\'eorica, Universidad Aut\'onoma de Madrid \& CSIC}
\centerline{\sl E-28049 Madrid, Spain}
\vskip0.5cm
\centerline{\sl $^b$ Department of Physics \& Helsinki Institute of Physics}
\centerline{\sl P.O. Box 64, FI-00014 University of Helsinki, Finland}
\vskip0.5cm
\centerline{\sl  $^b$ Institute for Theoretical Physics,  University of Regensburg,}    
 \centerline{\sl D-93040 Regensburg, Germany}
\vskip0.5cm
\begin{center}
{\sl  E-mail:} \hskip 1mm \href{mailto:marco.panero@inv.uam.es}{{\tt marco.panero@inv.uam.es}}, \href{mailto:kari.rummukainen@helsinki.fi}{{\tt kari.rummukainen@helsinki.fi}}, \href{mailto:andreas.schaefer@physik.uni-regensburg.de}{{\tt andreas.schaefer@physik.uni-regensburg.de}}
\end{center}
\vskip1.0cm
\begin{abstract}
We present a first-principle computation of the jet quenching parameter, which describes the momentum broadening of a high-energy parton moving through the deconfined state of QCD matter at high temperature. Following an idea originally proposed by Caron-Huot, we explain how one can evaluate the soft contribution to the collision kernel characterizing this real-time phenomenon, analyzing certain gauge-invariant operators in a dimensionally reduced effective theory (electrostatic QCD), which can be studied non-perturbatively via simulations on a Euclidean lattice. Our high-precision numerical computations at two different temperatures indicate that soft contributions to the jet quenching parameter are large. After discussing the systematic uncertainties involved, we present a quantitative estimate for the jet quenching parameter in the temperature range accessible at heavy-ion colliders, and compare it to results from phenomenological models as well as to strong-coupling computations based on the holographic correspondence.
\end{abstract}
\vspace*{0.2cm}
\noindent PACS numbers: 
12.38.Gc, %Lattice QCD calculations
12.38.Mh, %Quark-gluon plasma
11.10.Wx %Finite-temperature field theory

\end{titlepage}

\section{Introduction and motivation} 
\label{sec:intro}

Understanding strongly interacting matter under extreme conditions of temperature or density is one of the major challenges in high-energy physics~\cite{Fukushima:2011jc}. During the past few years, experiments carried out through heavy-ion collisions at SPS~\cite{Heinz:2000bk}, RHIC~\cite{Adcox:2004mh, Arsene:2004fa, Back:2004je, Adams:2005dq} and LHC~\cite{Aamodt:2010jd, Aamodt:2010cz, Aamodt:2010pb} confirmed the theoretical expectation that, at sufficiently high values of the temperature $T$ (around $160$~MeV~\cite{BraunMunzinger:2007zz}), hadrons turn into a qualitatively different, deconfined state of matter~\cite{Cabibbo:1975ig}: the ``quark-gluon plasma'' (QGP)~\cite{Shuryak:1978ij}. Important indirect evidence for the QGP includes the observation of elliptic flow~\cite{Ackermann:2000tr, Aamodt:2010pa, ATLAS:2011ah, ATLAS:2012at, Chatrchyan:2012ta}, characteristic photon and dilepton spectra~\cite{Adare:2008ab, Adare:2009qk}, quarkonium melting~\cite{Vogt:1999cu, Gerschel:1998zi, Adare:2006ns, Adare:2008sh, Aad:2010aa, Chatrchyan:2011pe, Chatrchyan:2012np, Abelev:2012rv}, strangeness enhancement~\cite{Andersen:1998vu, Andersen:1999ym, Antinori:2006ij, Abelev:2006cs, Abelev:2007rw, Abelev:2008zk} and jet quenching~\cite{Aggarwal:2001gn, Adcox:2001jp, Adler:2002xw, Adler:2002tq, Antinori:2005cx, Aad:2010bu, Chatrchyan:2011sx, Chatrchyan:2012gt} (see also refs.~\cite{Baier:2002tc, Stoecker:2004qu, Spousta:2013aaa, Veres:2013oga} for reviews).

Jet quenching (which, together with elliptic flow, can be considered as a ``gold-plated'' observable) was first discussed in refs.~\cite{Bjorken:1982qr, Bjorken:1982tu}. It consists in the suppression of hadrons of large transverse momentum, due to energy loss suffered by high-energy partons propagating through the deconfined medium, and belongs to the class of ``hard probes'' used in experimental investigations of the QGP. The phenomenological importance of jet quenching has motivated a large number of theoretical studies~\cite{Baier:1996sk, Baier:1996kr, Baier:2000mf, Bass:2008rv, Arnold:2008vd, Liu:2006ug,  CasalderreySolana:2011us, Wang:1991xy, Gyulassy:1993hr, Wang:1994fx, Zakharov:1996fv, Zakharov:1997uu, Zakharov:1998sv, Gyulassy:1999zd, Gyulassy:2000fs, Gyulassy:2000er, Wiedemann:2000za, Arnold:2001ba, Arnold:2001ms, Arnold:2002ja, Wang:2001ifa, Muller:2002fa, Kovner:2003zj, Armesto:2003jh, Jeon:2003gi, Salgado:2003gb, Gyulassy:2003mc, Zhang:2003yn, Armesto:2004pt, Vitev:2005yg, Koch:2005sx, Renk:2005si, Polosa:2006hb, Chiu:2006pu, Liu:2006he, Armesto:2006zv, Gubser:2006nz, Qin:2007rn, CasalderreySolana:2007qw, Majumder:2007hx, Majumder:2007ne, Gubser:2008as, Hatta:2008tx, Liu:2008tz, d'Enterria:2009am, Gubser:2009sn, Gursoy:2009kk, Qin:2009bk, Wiedemann:2009sh, Marquet:2009eq, Majumder:2010qh, Gursoy:2010aa, D'Eramo:2010ak, Armesto:2011ht, Kiritsis:2011bw, Beraudo:2011bh, Beraudo:2012bq, D'Eramo:2012jh, Liou:2013qya, Mehtar-Tani:2013pia, Shuryak:2013cja, Hidalgo-Duque:2013rta, Apolinario:2013foa, Burke:2013yra, Li:2014hja, Xu:2014ica, Iancu:2014kga, Wang:2014xda}, using perturbative weak-coupling expansions, various types of phenomenological modeling, or strong-coupling techniques based on the gauge/string duality~\cite{Maldacena:1997re, Gubser:1998bc, Witten:1998qj, Aharony:1999ti}. A quantity commonly used in the description of the phenomenon is the jet quenching parameter $\qhat$, which represents the average increase in the squared transverse component of parton momentum per unit length~\cite{Baier:1996sk, Baier:1996kr, Baier:2000mf}. Unfortunately, extracting $\qhat$ from experimental observables is quite non-trivial~\cite{Bass:2008rv, Renk:2010ac, Renk:2014nwa}, as the numerical values obtained depend on the details of the underlying dynamics. This observation can, however, be turned around: if an accurate, first-principle theoretical prediction for $\qhat$ could be worked out, e.g. from a lattice calculation, then different observables would provide valuable information on this dynamics, and thus help to better understand actual heavy-ion collisions.

Unfortunately, since jet quenching is a \emph{real-time} phenomenon, the possibility of investigating it by means of simulations on a \emph{Euclidean} lattice is not obvious at all.\footnote{For an overview of recent lattice studies of quantities relevant for QGP dynamical properties and hard and thermal probes, see ref.~\cite{Ding:2014xha}.} In this article, however, following a proposal originally put forward in ref.~\cite{CaronHuot:2008ni}, we show that a rigorous, systematic approach to the problem is possible. This enabled us to carry out a non-perturbative determination of the contributions to $\qhat$ from long-wavelength modes of the quark-gluon plasma. When combined with perturbative information relevant for the dynamics of hard thermal modes, these results lead to a model-independent quantitative estimate for the jet quenching parameter at temperatures accessible to experiments with present technology.

In section~\ref{sec:setup} we review the basic theoretical formalism to describe the jet quenching parameter $\qhat$ and introduce the formulation of a dimensionally reduced effective theory for hot QCD (electrostatic QCD, or EQCD). This effective theory captures the dynamics of long-wavelength modes in a non-Abelian plasma. Then, in section~\ref{sec:lattice} we present a non-perturbative study of $\qhat$ via lattice simulations of EQCD. The key point, making it possible to extract information relevant for a parameter describing a dynamical, real-time phenomenon via simulations in a Euclidean setup, is that the soft physics contributions of interest manifest themselves in the form of correlation functions near the light cone, which can be exactly mapped to Euclidean correlators (see refs.~\cite{CaronHuot:2008ni, Ghiglieri:2013gia, Cherednikov:2013pba} for a discussion). An intuitive physical argument can be summarized as follows: the contribution to the collisional rate from plasma modes collinear with the jet is suppressed by a reduced center-of-mass energy~\cite{Braaten:1991jj, Braaten:1991we}, so the contribution to the collision kernel $C(p_\perp)$~describing the evolution of jets---see eq.~(\ref{qhat_definition}) below---from soft (and essentially classical) modes is not sensitive to the precise value of the velocity of the ultrarelativistic parton moving through the plasma. This implies that $C(p_\perp)$ would be essentially unchanged, even if the parton moved at a speed larger than $1$ (in natural units). Although this case is not physical, it would correspond to making the trajectory of the hard parton spacelike, opening up the possibility of carrying out its computation in a Euclidean setup.\footnote{Other recent works discuss how to extract light-cone physics from simulations on a Euclidean lattice using a different approach~\cite{Musch:2011er, Ji:2013dva, Lin_Lattice2013, Lin:2014zya}.}

More precisely, the kernel describing the parton-plasma constituents collisions is related to two-point correlators of spatially separated (assume $|t|<|z|$) light-like Wilson lines. In four-momentum space, the latter can be re-written in terms of the difference between a retarded (R) and an advanced (A) Green's function:
\begin{eqnarray}
\label{Minkowski_to_Euclidean_derivation_1}
&& G^<(t,x_\perp,z) = \int \dd \omega \int \dd^2 p_\perp \int \dd p^z \tilde{G}^<(\omega,p_\perp,p^z) e^{-i \left( \omega t - x_\perp \cdot p_\perp - z p^z \right)} \nonumber \\
&& \;\;\;= \int \dd \omega \int \dd^2 p_\perp \int \dd p^z \left[ \nb (\omega) + \frac{1}{2} \right] \left[ \tilde{G}_{\mbox{\tiny{R}}}(\omega,p_\perp,p^z) - \tilde{G}_{\mbox{\tiny{A}}}(\omega,p_\perp,p^z) \right] e^{-i \left( \omega t - x_\perp \cdot p_\perp - z p^z \right)}, \qquad
\end{eqnarray}
where $\nb(\omega) = 1/[\exp(\omega / T)-1]$ denotes the Bose-Einstein distribution. Shifting the momentum component along the direction of motion as $p^{\prime z}=p^z - \omega t/z$, the integration over frequencies in eq.~(\ref{Minkowski_to_Euclidean_derivation_1}) becomes trivial: for the term involving the  retarded (advanced) Green's function, it can be carried out by analytical continuation into the upper (lower) complex frequency semi-plane, without encountering any singularities, except those of the function $[\nb(\omega) + 1/2]$. Since the latter has simple poles at $\omega=2 \pi i n T$ for integer $n$ (with residues $T$), the previous expression can be re-written as a sum over Matsubara frequencies,
\begin{equation}
\label{Minkowski_to_Euclidean_derivation_2}
G^<(t,x_\perp,z) = T \sum_{n \in \Z } \int \dd^2 p_\perp \int \dd p^{\prime z} \tilde{G}_{\mbox{\tiny{E}}}(2 \pi n T,p_\perp,p^{\prime z}+ 2 \pi i n T t/z) e^{i \left( x_\perp \cdot p_\perp + z p^{\prime z} \right)} ,
\end{equation}
where $\tilde{G}_{\mbox{\tiny{E}}}$ is the Euclidean correlator, related to the retarded Green's function via $\tilde{G}_{\mbox{\tiny{E}}}(\omega,p_\perp,p^z)=i\tilde{G}_{\mbox{\tiny{R}}}(-i\omega,p_\perp,p^z)$. Note that contributions from $n \neq 0$ modes are exponentially suppressed at large separations, while the $gT$ contribution comes only from the zero mode, which is time-independent, and can be evaluated in EQCD.

We remark that, while the results of our simulations in the dimensionally reduced effective theory do not include \emph{all} contributions to $\qhat$ (by construction, this effective theory misses contributions from the ``hard'' scale $\pi T$), they do capture the contributions responsible for large soft corrections. The perturbative analysis of the latter cannot be pushed beyond the next-to-leading order~\cite{CaronHuot:2008ni}, since at subsequent orders it becomes intrinsically non-perturbative. In this article we will use the $\qhatsoft$ notation to denote contributions to the jet quenching parameter that are captured within the EQCD framework.

In section~\ref{sec:discussion} we discuss our results, comparing them with other recent, related studies~\cite{Laine:2012ht, Benzke:2012sz, Laine:2013lia, Laine:2013apa}, and list future research directions. For a concise summary of the present work, see also refs.~\cite{Panero:2013fva, Panero:2013rra}.

\section{Theoretical setup}
\label{sec:setup}

Describing the jet quenching phenomenon theoretically means deriving the dynamical effects experienced by a hard parton moving through the quark-gluon plasma. The standard formalism reviewed in ref.~\cite{CasalderreySolana:2007zz} models the quark-gluon plasma as a sort of ``background field'', neglecting the back-reaction caused by the hard parton propagation. In the eikonal approximation~\cite{Kovner:2001vi}, the $S$-matrix~elements relating initial and final states of a parton subject to multiple scatterings off medium constituents can be written in terms of a Wilson line along a light-cone trajectory
\begin{equation}
\label{S_matrix_element}
\mathcal{W}(x) = \mathcal{P} \exp \left[ ig \int \dd x_+ A_-(x_+,x_\perp) \right],
\end{equation}
where $x_\pm$ denote light-cone coordinates
\begin{equation}
\label{lc_coordinates}
x_{\pm}=\frac{t \pm z}{\sqrt{2}}
\end{equation}
and $x_\perp$ represents the transverse coordinates. If the parton energy is much larger than the typical energy of QGP excitations, the leading effect experienced by the hard parton is an increase of the modulus of its transverse momentum component $p_\perp$. More quantitatively, one can show that the square of the transverse momentum component and the energy loss of the parton depend linearly on the distance $L$ travelled through the medium~\cite{Baier:1996sk, Baier:1996kr, Baier:2000mf}. This leads to define the jet quenching parameter $\qhat$ as
\begin{equation}
\label{qhat_definition}
\qhat = \frac{\langle p^2_\perp \rangle}{L} = \int \frac{ \dd^2 p_\perp}{(2\pi)^2} p^2_\perp C(p_\perp) ,
\end{equation}
where $C(p_\perp)$ represents the ``transverse collision kernel''. $C(p_\perp)$ is nothing but the differential collision rate describing the interaction between the hard parton (in the eikonal approximation) and the plasma constituents.

In particular, for the case of a light quark probe and for transverse momenta much smaller than the hard scale, $p_\perp \ll T$, the perturbation theory (PT) analysis carried out at leading order (LO) in ref.~\cite{Arnold:2008vd} and extended to next-to-leading order (NLO) in ref.~\cite{CaronHuot:2008ni} yields
\begin{eqnarray}
\label{perturbative_C_soft}
C(p_\perp) &=& g^2 T \Cfund \left\{ \frac{1}{p_\perp^2}-\frac{1}{p_\perp^2+\mE^2}\right\} + g^4 T^2 \Cfund \Cadj \left\{ \frac{7}{32 p_\perp^3 } - \frac{3 \mE p_\perp + 2(p_\perp^2-\mE^2)\arctan (p_\perp/\mE) }{4 \pi p_\perp (p_\perp^2+\mE^2)^2} \right. \nonumber \\
&& \left. + \frac{2\mE p_\perp - (p_\perp^2+4\mE^2)\arctan [p_\perp/(2\mE)]}{16 \pi p_\perp^5} - \frac{\arctan (p_\perp/\mE)}{2 \pi p_\perp (p_\perp^2+\mE^2)} \right. \nonumber \\
&& \left.  + \frac{\arctan [p_\perp/(2\mE)]}{2 \pi p_\perp^3} + \frac{\mE}{4 \pi (p_\perp^2+\mE^2)} \left( \frac{3}{p_\perp^2+4\mE^2} -\frac{1}{p_\perp^2} \right)  \right\} + O(g^6),
\end{eqnarray}
where $\Cfund=(N^2-1)/(2N)$ and $\Cadj=N$ are, respectively, the eigenvalues of the quadratic Casimir operator for the fundamental and for the adjoint $\SU(N)$ representation, $N=3$ denotes the number of color charges, while $\mE = gT\sqrt{(\Cadj + n_f \tfund)/3}$ is the Debye mass parameter, with $\tfund=1/2$ the trace normalization for the fundamental representation and $n_f$ the number of dynamical light quark flavors. For momenta $p_\perp \sim gT$, eq.~(\ref{perturbative_C_soft}) gives the full $O(g)$ correction to the collision kernel. On the other hand, for momenta $p_\perp \gg gT$ one obtains~\cite{Arnold:2008vd}
\begin{eqnarray}
\label{perturbative_C_hard}
C(p_\perp) = \frac{g^4 T^3 \Cfund}{\pi^2 p_\perp^4} \left[ 2 \Cadj I_+\left(p_\perp/T\right) + 4 n_f \tfund I_-\left(p_\perp/T\right) \right] + O(g^6) ,
\end{eqnarray}
with
\begin{equation}
\label{I_functions}
I_\pm(x) = x^2 \sum_{a=1}^{\infty} \sum_{b=0}^{\infty} (\pm 1)^{a+b-1} \frac{ab}{2(a+b)^3} K_2 \left( x \sqrt{ab} \right) ,
\end{equation}
where $K_2(z)$ is a modified Bessel function of the second kind.

The integral on the right-hand side of eq.~(\ref{qhat_definition}) should be restricted to transverse momenta of modulus not larger than $\qmax$, where the definition of the upper bound $\qmax$ (and, therefore, of $\qhat$) is ambiguous.\footnote{Note that, if the jet quenching parameter~is computed introducing an auxiliary scale $gT \ll q^\star \ll \pi T$ to separate the regimes in which the ``soft'' and ``hard'' expressions for the collision kernel hold, one may also expect $\qhat$ to depend~on $q^\star$. This turns out not to be the case: the $q^\star$-dependent contributions to $\qhat$ from eq.~(\ref{perturbative_C_soft}) and eq.~(\ref{perturbative_C_hard}) cancel against each other~\cite{CaronHuot:2008ni, Laine:2012ht}.} This restriction is necessary because at large $p_\perp$ the leading-order contribution to $C(p_\perp)$ is proportional to $1/p_\perp^4$ (see also ref.~\cite{Aurenche:2002pd} for a discussion): this implies that $\qhat$ depends \emph{logarithmically} on $\qmax$. 

Starting from the definition eq.~(\ref{qhat_definition}), $\qhat$ can be derived from expectation values of a pair of light-cone Wilson lines of length $\ell$, with the inclusion of two transverse lines of length $r$, to enforce gauge invariance. From the expectation value of the trace of the resulting Wilson loop $ W(\ell,r) $, one can then extract the quantity $V(r)$ defined as
\begin{equation}
\label{potential}
V(r) = - \lim_{\ell \to \infty} \frac{1}{\ell} \ln \langle W(\ell,r) \rangle,
\end{equation}
which equals minus the Fourier transform of $C(p_\perp)$ (up to subtraction of a divergent contribution):
\begin{equation}
\label{potential_and_C}
V(r) = \int \frac{\dd^2 p_\perp}{(2 \pi)^2} \left[ 1 - \exp \left( i x_\perp \cdot p_\perp \right) \right] C(p_\perp),
\end{equation}
with $ r = |x_\perp| $.

As we mentioned above, a LO perturbative evaluation of the transverse collision kernel was carried out in ref.~\cite{Arnold:2008vd}, and later extended to NLO in ref.~\cite{CaronHuot:2008ni}. However, it is well known that perturbative expansions in non-Abelian gauge theories at finite temperature are hampered by serious convergence problems (see, e.g.~\cite{Kajantie:2002wa}, and references therein, for a discussion about the case of the free energy) and by a mathematically non-trivial structure---involving, in particular, terms which are not proportional to integer powers of $\alpha_s$. This can be traced back to the existence of infrared divergences~\cite{Linde:1980ts, Gross:1980br}, which lead to a breakdown of the familiar (at zero temperature) correspondence between expansions in powers of $\alpha_s$ and expansions in number of loops. A consistent way to deal with these effects is based on the introduction of a suitable hierarchy of effective theories, defined in three spatial dimensions, relying on the separation between ``hard'', $O(\pi T)$, ``soft'', $O(gT)$, and ``ultrasoft'', $O(g^2T/\pi)$ energy scales~\cite{Ginsparg:1980ef, Appelquist:1981vg, Kajantie:1995dw, Braaten:1989kk, Braaten:1995cm, Braaten:1995jr}---see also refs.~\cite{Braaten:1989mz, Frenkel:1989br, Braaten:1990az, Braaten:1991gm, Blaizot:2001nr}. With an appropriate definition of their parameters (which can be fixed by a \emph{matching} procedure, imposing consistency with the original theory), these effective theories capture and properly resum the physics of long-wavelength modes, leaving a well-behaved perturbative series behind.

To be definite, the Euclidean formulation of equilibrium, finite-temperature QCD in four dimensions (4D) is based on the Lagrangian
\begin{equation}
\label{QCD_Lagrangian}
\LQCD = \frac{1}{4} F_{\mu \nu}^a F_{\mu \nu}^a + \sum_{f=1}^{n_f} \bar{\psi}_f \left[ \gamma_\mu (\partial_\mu + i g A_\mu^a t^a)+ m_f \right] \psi_f ,
\end{equation}
where $g$ is the (bare) gauge coupling, $t^a$ are the Hermitian generators of $\SU(3)$, the field-strength components are defined as $F_{\mu \nu}^a = \partial_\mu A_\nu^a - \partial_\nu A_\mu^a - g f^{abc} A_\mu^b A_\nu^c$, and the Dirac operator involves a summation over $n_f$ quark flavors, of (generically different) bare masses $m_f$. In the following, we restrict our attention to $n_f=0$, $2$ or $3$ light flavors, assuming that their masses are much smaller than the system temperature: $m_f/T \ll 1$. The partition function is obtained by functional integration over bosonic (fermionic) fields, with (anti-)periodic boundary conditions along the compactified Euclidean time direction, whose extent is the inverse of the physical temperature $T$. At high temperatures, the theory is in a deconfined phase. Due to asymptotic freedom, the renormalized physical coupling runs logarithmically with the energy scale, so that, in particular, perturbation theory becomes reliable to describe the physics of ``hard'' processes.

Starting from eq.~(\ref{QCD_Lagrangian}), at sufficiently high temperatures one can construct an effective theory by integrating out all non-zero momentum modes for the Euclidean-time components of the gauge fields:\footnote{Note that, owing to the anti-periodic boundary conditions along the Euclidean time direction, the minimum Matsubara frequency for fermion fields is finite, and proportional to $T$.} this results in a three-dimensional (3D) effective theory, called electrostatic QCD (EQCD), which is described by the Lagrangian
\begin{equation}
\label{EQCD_Lagrangian}
\mathcal{L} = \frac{1}{4} F_{ij}^a F_{ij}^a + \tr \left( (D_i A_0)^2 \right) + \mE^2 \tr \left( A_0^2 \right) + \lambda_3 \left( \tr \left( A_0^2 \right) \right)^2 ,
\end{equation}
with $D_i=\partial_i + i \gE [A_i, \cdot ]$, i.e. 3D $\SU(3)$ Yang-Mills theory coupled to an adjoint Higgs field $A_0$. In three dimensions the $A_i$ (and $A_0$) fields, as well as the $\gE$~gauge coupling, have energy dimension equal to one half, while the coupling of the quartic self-interaction term $\lambda_3$ has dimensions of an energy (like the $\mE$ mass parameter). The theory is super-renormalizable, and, in the modified minimal-subtraction scheme, the physical squared mass of the scalar field, evaluated at a generic momentum scale $\mu$, reads
\begin{equation}
\label{running_mass}
\mE^2(\mu) = \frac{5\lambda_3}{8 \pi^2} (\lambda_3 - 3 \gE^2) \ln \left( \frac{\mu^2}{\Lambda_0^2} \right)
\end{equation}
(with $\Lambda_0$ a constant momentum scale characterizing the theory), while $\gE$ and $\lambda_3$ turn out to be renormalization-scale independent at leading order. 

Note that the most general Lagrangian of a three-dimensional super-renormalizable $\SU(3)$-plus-Higgs theory could also include $\tr (A_0^3)$ and $\tr (A_0^2) \tr (A_0^3)$ terms. However, operators odd in $A_0$ appear in EQCD only in the presence of quark chemical potential, and thus in the following we do not consider these. Note also that the Lagrangian in eq.~(\ref{EQCD_Lagrangian}) does not contain a $\tr \left( A_0^4 \right)$ term, because for fields in the algebra of $\SU(3)$---as well as $\SU(2)$---generators, such operator is trivially related to the quartic term already included in eq.~(\ref{EQCD_Lagrangian}) by the $2\tr \left( A_0^4 \right) = \left[ \tr \left( A_0^2 \right) \right]^2$ identity.\footnote{For the $\SU(2)$ theory, this identity can be easily proven noting that (in the fundamental representation and with the normalization specified above) the generators satisfy: $T^a T^b = 2i \epsilon^{abc} T^c + \delta^{ab} \ide$. For the $\SU(3)$ theory, the same identity is a consequence of the fact that: $3 \{ T^a, T^b\} = \delta^{ab} \ide +3 d^{abc} T^c$, where the totally symmetric $d^{abc}$ tensor satisfies: $ 3 ( d^{abc} d^{egc} + d^{aec} d^{bgc} + d^{agc} d^{ebc} ) = \delta^{ab} \delta^{eg} + \delta^{ae} \delta^{bg} + \delta^{ag} \delta^{eb}$.} Note, however, that this identity is specific of the $\SU(2)$ and $\SU(3)$ generators, and does not hold for $\SU(N > 3)$.

Since the Lagrangian in eq.~(\ref{EQCD_Lagrangian}) does not include odd powers of $A_0$, it is classically invariant under a $\Z_2$ symmetry which flips the sign of this field. At the quantum level, the ground state of the theory may be invariant under or spontaneously break this discrete symmetry. Only the symmetric phase, however, corresponds to high-temperature QCD.

In order to reproduce the soft physics of high-temperature QCD, all parameters of the theory described by eq.~(\ref{EQCD_Lagrangian}), namely $\gE$, $\mE$ and $\lambda_3$, have to be appropriately fixed via matching to the theory described by eq.~(\ref{QCD_Lagrangian})~\cite{Nadkarni:1982kb, Nadkarni:1988fh, Landsman:1989be, Kajantie:1997pd, Kajantie:1997tt}.\footnote{Recently, a similar approach has also been discussed for the dimensional reduction of a model with a large extra-dimension~\cite{deForcrand:2010be}.} In particular, a weak-coupling analysis shows that the leading-order relation between the 3D theory parameter $\gE$ and the running coupling and temperature of the 4D theory (at a given momentum scale $\mu$) already arises at tree level:
\begin{equation}
\gE^2=g^2(\mu)T  .
\end{equation}
By contrast, the relations for $\mE$ and for $\lambda_3$ can be derived from one-loop computations: as we already mentioned above, the $\mE$ mass corresponds to the Debye mass parameter, associated with screening of chromoelectric modes in the QCD plasma,
\begin{equation}
\mE^2 = \left( 1 + \frac{n_f}{6} \right) g^2(\mu)T^2  ,
\end{equation}
while
\begin{equation}
\lambda_3 = \frac{9- n_f}{24 \pi^2}g^4(\mu)T  .
\end{equation}
An optimal value for the momentum scale $\mu$ can be defined, by extending the computation to two-loop order~\cite{Kajantie:1997tt}. It is convenient to introduce dimensionless ratios
\begin{equation}
x = \frac{\lambda_3}{\gE^2}, \qquad \qquad y = \frac{\mE^2(\mu=g^2_3)}{\gE^4}  .
\end{equation}
With this parametrization, the 3D theory describes soft physics of QCD with $n_f$ light quark flavors when $x$ and $y$ satisfy the relation
\begin{equation}
y(x) = \frac{(6+n_f)(9-n_f)}{144 \pi^2 x} + \frac{486 -33 n_f -11 n_f^2 -2 n_f^3}{96(9-n_f) \pi^2 } + O(x).
\end{equation}
In particular, $y$ is related to the temperature of the four-dimensional theory via
\begin{equation}
\frac{T}{\LambdaMSbar} \simeq \left\{ 
\begin{array}{ll}
\exp( 7.2 y - 2.1) \qquad & \mbox{for pure $\SU(3)$ Yang-Mills theory} \\
\exp( 6.1 y - 2.3) \qquad & \mbox{for QCD with $n_f=2$ light flavors} \\
\exp( 5.9 y - 2.4) \qquad & \mbox{for QCD with $n_f=3$ light flavors}
\end{array}
\right.  ,
\end{equation}
where $\LambdaMSbar$ is the dynamically generated QCD scale in the modified minimal-subtraction scheme. The latter quantity appears to be only mildly dependent on the number of flavors~\cite{Gockeler:2005rv} and its value for $n_f=2$ is $\LambdaMSbar = 255^{+40}_{-15}$~MeV~\cite{Kneur:2011vi}.

An important aspect concerns the r\^ole of center symmetry. The Lagrangian of $\SU(N)$~Yang-Mills theory ($n_f=0$) in four spacetime dimensions is characterized by a global $\Z_N$ symmetry, which gets spontaneously broken in the deconfined phase.\footnote{By contrast, in QCD with $n_f \neq 0$ dynamical quark flavors the center symmetry is not exact, being explicitly broken by the Dirac operator.} In fact, at high temperature, there exist $N$ different ground states, which can be labelled by the Polyakov loop expectation value. In particular, in all of these phases but one, $A_0$ takes large values, of order $T/g$. Since the dimensionally reduced effective theory only describes an adjoint field of small amplitude, it does not capture the $\Z_N$ symmetry\footnote{A center-symmetric dimensionally reduced effective theory, however, has been proposed in ref.~\cite{Vuorinen:2006nz} (for $N=3$) and in ref.~\cite{deForcrand:2008aw} (for $N=2$).}---at least, not fully.\footnote{A remnant of $\Z_N$ symmetry is, nevertheless, preserved in the 3D theory, which, on its critical curve, exhibits $N$ separate (but physically not equivalent) metastable states.}

For comparison with previous studies, in sect.~\ref{sec:lattice} we will also present results for Wilson loops in magnetostatic QCD (MQCD): this is a dimensionally reduced effective theory for hot QCD, which describes the static chromomagnetic sector at scales of order $ g^2 T/\pi$, and is just three-dimensional pure Yang-Mills theory. It can be formally obtained from EQCD, by integrating out the $g T$ scale---which includes, in particular, the physics of the $A_0$ field. The relation between the gauge couplings in MQCD and in EQCD is of the form $\gsqM = \gsqE [1 + O(\gsqE)]$~\cite{Giovannangeli:2003ti}. There exists convincing numerical evidence that the square root of the spatial string tension describing the area-law decay of large spatial Wilson loops in the deconfined phase of the theory in $(3+1)$ dimensions scales well with $\gsqM$, and is consistent with the string tension of MQCD; this already holds for temperatures equal to twice (or even one and a half times) the deconfinement temperature~\cite{Boyd:1996bx, Bali:1993tz, Bali:1993ub, Cheng:2008bs}.

\section{Lattice study}
\label{sec:lattice}

The lattice study presented here was performed using the Wilson discretization of the continuum EQCD action, which is obtained integrating eq.~(\ref{EQCD_Lagrangian}) over 3D space:
\begin{eqnarray}
\label{lattice_action}
&& S = \beta \sum_{x \in {\boldsymbol\Lambda}} \sum_{1 \le k < l \le 3} \left( 1 -\frac{1}{3} \real \tr
U_{k,l}(x) \right) - \frac{12}{\beta} \sum_{x \in {\boldsymbol\Lambda}} \sum_{1 \le k \le 3} \tr \left(
A_0(x) U_k(x) A_0(x+a\hat{{k}}) U_k^\dagger(x) \right) \nonumber \\
&& \qquad + \sum_{x \in {\boldsymbol\Lambda}} \left[ \alpha \tr A_0^2(x)  + \left(\frac{6}{\beta}\right)^3 x_{\mbox{\tiny{L}}} \left( \tr A_0^2(x) \right)^2 \right]  ,
\end{eqnarray}
where all summations over $x$ run over the sites of a three-dimensional isotropic cubic lattice ${\boldsymbol\Lambda}$ of spacing $a$ and physical volume $V=a^3 N_x N_y N_z$. Here $A_0(x)$ denotes the adjoint lattice field at a generic site $x$, $U_k(x)$ is a lattice link variable (taking values in the $\SU(3)$ group), while
\begin{equation}
\label{plaquette}
U_{k,l}(x)=U_{k}(x)U_{l}( x+a\hat{k} ) U_{k}^\dagger ( x+a\hat{l} )U_{l}^\dagger(x)
\end{equation}
is the plaquette in the $(k,l)$ plane through a generic site $x$, and, following notations almost identical to those of ref.~\cite{Hietanen:2008tv}, we denoted
\begin{eqnarray}
&& \beta = \frac{6}{\gE^2 a}, \qquad x = \frac{\lambda_3}{\gE^2}, \qquad x_{\mbox{\tiny{L}}} =  x + \frac{0.328432-0.835282x+1.167759 x^2}{\beta}, \nonumber \\
&& y_{\mbox{\tiny{L}}} =  y = \frac{\mE^2(\bar{\mu}=g^2_3)}{\gE^4}, \nonumber \\
&& \alpha = \frac{6^2}{\beta} \left\{  1 + \frac{6}{\beta^2} y_{\mbox{\tiny{L}}}
 - ( 3 + 5 x_{\mbox{\tiny{L}}} )\frac{3.175911535625}{ 2\pi\beta} \right.\nonumber \\
&& \qquad \left. - \frac{3}{2 (2\pi\beta)^2} \left[ 20x_{\mbox{\tiny{L}}} (3 -x_{\mbox{\tiny{L}}})(\ln\beta+0.08849)+ 34.768x_{\mbox{\tiny{L}}}+36.130 \right] \right\}  .
\end{eqnarray}
These relations already account for $O(1/a)$ and $O(\ln a)$ divergences associated with the lattice ``mass term'' $\alpha$. Since the theory is super-renormalizable, there are no other divergences. The relations are improved to $O(a)$, excluding an additive $O(a)$ contribution to the mass term.

Note that, according to our conventions, the (tree-level) relation between lattice and continuum fields involves a proportionality factor $\gE$, e.g. $A_i^{\mbox{\tiny{(cont)}}}(x) = \gE A_i^{\mbox{\tiny{(latt)}}}(x)$, so that lattice fields are dimensionless. In order not to encumber  our notations, in the rest of this section we will, however, omit the ``latt'' superscript.

The phase structure of the lattice theory described by eq.~(\ref{lattice_action}) includes a symmetric phase, in which the expectation value of $\tr A_0^2(x)$ is small, and a broken-symmetry phase~\cite{Kajantie:1997tt, Kajantie:1998yc}. Our simulations are performed in the symmetric phase, which is the one describing the physics of soft modes in high-temperature QCD. Owing to the finiteness of the mass gap, finite-volume effects are exponentially suppressed with the product of the linear size of the system and the mass of the lightest physical state in the spectrum. In practice, finite-volume effects are completely negligible for all $ \beta < N_s $ (where $N_s$ denotes the minimum among $N_x$, $N_y$ and $N_z$). The parameters of our simulations, summarized in tab.~\ref{tab:parameters}, satisfy this condition.

\begin{table}
\begin{center}
\begin{tabular}{|c|c|c|c|c|c|c|}
\hline
 $\beta$ & $x_{\mbox{\tiny{L}}}$ & $y_{\mbox{\tiny{L}}}$ & $N_x$ & $N_y$ & $N_z$ & total statistics \\
\hline
 12 & 0.1  & 0.448306  &  24 &  24 &  48 & $(7.07 \times 10^5) \times 10$ \\
 12 & 0.06 & 0.710991  &  24 &  24 &  48 & $(7.07 \times 10^5) \times 10$ \\
 12 & ---  & ---       &  24 &  24 &  48 &  $(1.4 \times 10^6) \times 10$ \\
 14 & 0.1  & 0.448306  &  24 &  24 &  48 & $(7.07 \times 10^5) \times 10$ \\
 14 & 0.06 & 0.710991  &  24 &  24 &  48 & $(7.07 \times 10^5) \times 10$ \\
 14 & ---  & ---       &  24 &  24 &  48 &  $(1.4 \times 10^6) \times 10$ \\
 16 & 0.1  & 0.448306  &  24 &  24 &  48 & $(7.07 \times 10^5) \times 10$ \\
 16 & 0.06 & 0.710991  &  24 &  24 &  48 & $(7.07 \times 10^5) \times 10$ \\
 16 & ---  & ---       &  24 &  24 &  48 &  $(1.4 \times 10^6) \times 10$ \\
 18 & 0.1  & 0.448306  &  24 &  24 &  48 & $(7.07 \times 10^5) \times 10$ \\
 18 & 0.06 & 0.710991  &  24 &  24 &  48 & $(7.07 \times 10^5) \times 10$ \\
 18 & ---  & ---       &  24 &  24 &  48 &  $(1.4 \times 10^4) \times 10$ \\
 24 & 0.1  & 0.448306  &  48 &  48 &  96 &    $(7 \times 10^3) \times 10$ \\
 24 & 0.06 & 0.710991  &  48 &  48 &  96 &    $(7 \times 10^3) \times 10$ \\
 24 & ---  & ---       &  48 &  48 &  96 &  $(1.4 \times 10^4) \times 10$ \\
 32 & 0.1  & 0.448306  &  64 &  64 & 132 &    $(7 \times 10^3) \times 10$ \\
 32 & 0.06 & 0.710991  &  64 &  64 & 132 &    $(7 \times 10^3) \times 10$ \\
 32 & ---  & ---       &  64 &  64 & 132 &  $(1.4 \times 10^4) \times 10$ \\
 40 & 0.1  & 0.448306  &  80 &  80 & 168 &    $(7 \times 10^3) \times 10$ \\
 40 & 0.06 & 0.710991  &  80 &  80 & 168 &    $(7 \times 10^3) \times 10$ \\
 40 & ---  & ---       &  80 &  80 & 168 &  $(1.4 \times 10^4) \times 10$ \\
 54 & 0.1  & 0.448306  &  80 &  80 & 168 &    $(7 \times 10^3) \times 10$ \\
 54 & 0.06 & 0.710991  &  80 &  80 & 168 &    $(7 \times 10^3) \times 10$ \\
 54 & ---  & ---       &  80 &  80 & 168 &  $(1.4 \times 10^4) \times 10$ \\
 67 & 0.1  & 0.448306  & 100 & 100 & 168 &    $(7 \times 10^3) \times 10$ \\
 67 & 0.06 & 0.710991  & 100 & 100 & 168 &    $(7 \times 10^3) \times 10$ \\
 67 & ---  & ---       & 100 & 100 & 168 &  $(1.4 \times 10^4) \times 10$ \\
 80 & 0.1  & 0.448306  & 120 & 120 & 168 &    $(7 \times 10^3) \times 10$ \\
 80 & 0.06 & 0.710991  & 120 & 120 & 168 &    $(7 \times 10^3) \times 10$ \\
 80 & ---  & ---       & 120 & 120 & 168 &  $(1.4 \times 10^4) \times 10$ \\
\hline
\end{tabular}
\end{center}
\caption{Parameters of our simulations, as defined in the text. The figures shown for the total statistics indicate that we performed independent simulations for each value of the parameters $\beta$, $x_{\mbox{\tiny{L}}}$, $y_{\mbox{\tiny{L}}}$ \emph{and} $n_r$ (for all integer values of $n_r$ from 1 to 10), as defined in the text. EQCD simulations at $x_{\mbox{\tiny{L}}}=0.1$ correspond to our lower temperature ensemble, while those at at $x_{\mbox{\tiny{L}}}=0.06$ correspond to the higher temperature. The parameters of pure Yang-Mills simulations, corresponding to MQCD ensembles, are listed in the table entries where the $x_{\mbox{\tiny{L}}}$ and $y_{\mbox{\tiny{L}}}$ values are not indicated.}
\label{tab:parameters}
\end{table}

\begin{figure*}
\centering
\includegraphics[width=.7\textwidth]{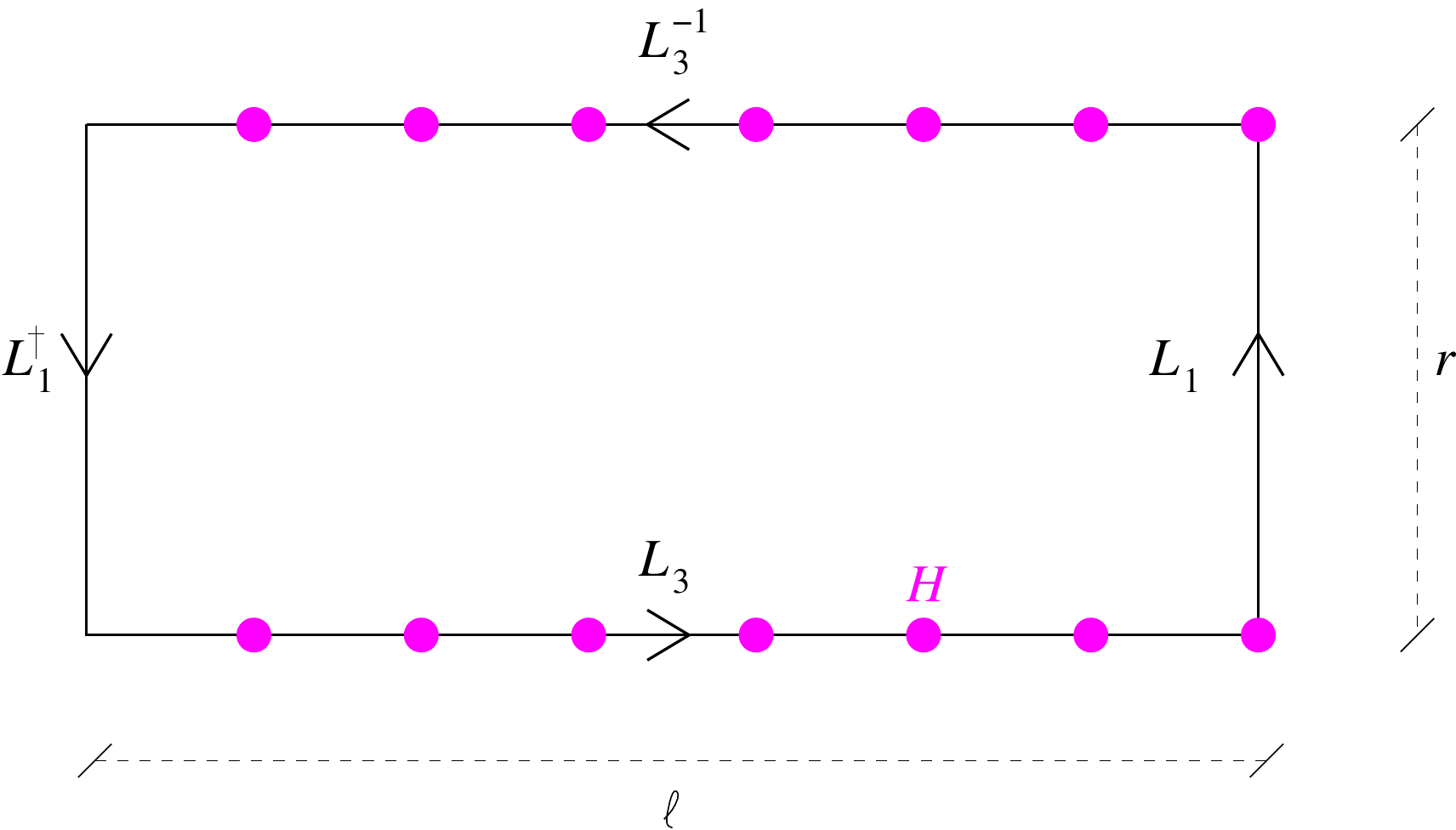}
\caption{In our EQCD simulations, the coordinate-space transverse collision kernel $V(r)$ is extracted from gauge-invariant operators defined in eq.~(\protect\ref{decorated_loop}) and given by the trace of ``decorated'' Wilson loops, obtained by dimensional reduction of Wilson loops parallel to a light-cone direction in the original theory. The figure shows a cartoon representing one of these operators, with the magenta circles denoting the insertion of parallel transporters obtained by exponentiation of the scalar field $A_0$ according to eq.~(\protect\ref{H_definition}).}
\label{fig:decorated_loop}
\end{figure*}

To obtain the transverse collisional kernel in coordinate space $V(r)$, we computed expectation values of traces of Wilson loops $W$ with two long ``decorated'' sides---see fig.~\ref{fig:decorated_loop}. They are lattice operators representing the counterpart (in the dimensionally reduced effective theory) of long Wilson loops stretching along a light-cone direction (in the original, four-dimensional theory). If the spatial projection of the light-cone Wilson line lies along direction $\hat{3}$, and the two light-cone Wilson lines are spatially separated along direction $\hat{1}$, then these lattice operators can be defined as
\begin{equation}
\label{decorated_loop}
W(\ell,r) = \tr \left( L_3(x,a n_\ell ) L_1 ( x + a n_\ell \hat{3}, a n_r ) L^{-1}_3(x+a n_r\hat{1}, a n_\ell ) L^\dagger_1(x,an_r)\right) ,
\end{equation}
with
\begin{eqnarray}
&& L_3( x, a n_\ell ) = \prod_{n=0}^{n_\ell -1} U_3\left( x + a n \hat{3}\right) H \left( x + a (n+1) \hat{3} \right), \\
&& L_1( x, a n_r ) = \prod_{n=0}^{n_r-1} U_1\left( x + a n \hat{1}\right) ,
\end{eqnarray}
where $n_\ell = \ell/a$, $n_r = r/a$, and $H(x)$ is a Hermitian matrix obtained by exponentiation of $A_0(x)$, \emph{videlicet}
\begin{equation}
\label{H_definition}
H(x)=\exp[- a \gE^2 A_0(x)] .
\end{equation}
Note that $H(x)$ represents a parallel transporter along a \emph{real-time} interval of length $a$.

Our numerical evaluation of expectation values of the operator defined in eq.~(\ref{decorated_loop}) is based on an implementation of the multi-level algorithm~\cite{Luscher:2001up}, which allows exponential enhancement of the signal-to-noise ratio for long loops.

\begin{figure*}
\centering
\includegraphics[width=1.\textwidth]{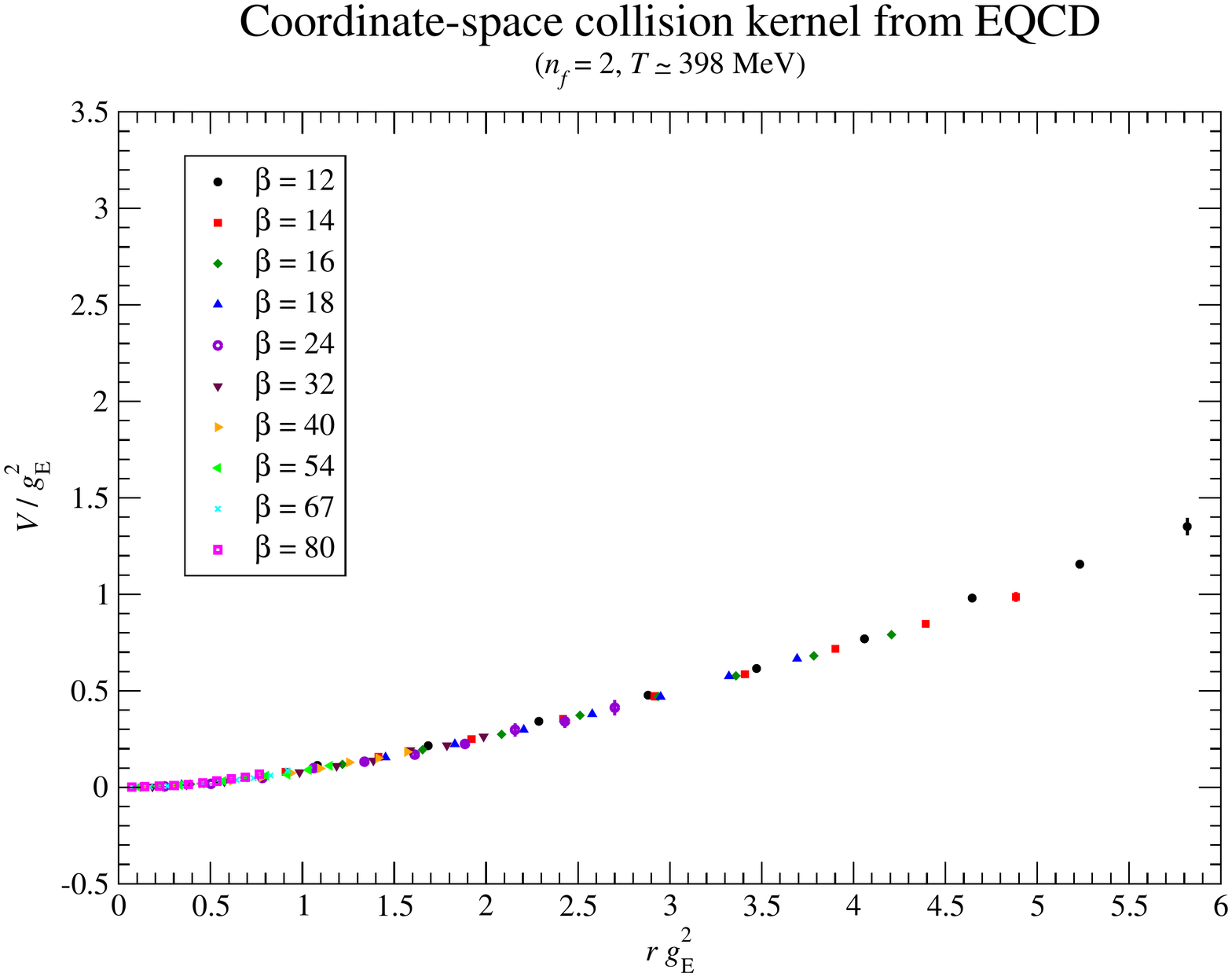}
\caption{Results of our EQCD simulations (at finite lattice cutoff and for a set of parameters corresponding to QCD with $n_f=2$ light quark flavors, at a temperature about $398$~MeV) for the collision kernel $V$, evaluated using eq.~(\protect\ref{potential}), as a function of the transverse loop size $r$. Both quantities are shown in appropriate units of $\gsqE$. Symbols of different colors denote results obtained from simulations at different lattice spacings, i.e. at different values of $\beta=6/(a \gsqE)$.}
\label{fig:colder_decorated_EQCD_V}
\end{figure*}

\begin{figure*}
\centering
\includegraphics[width=1.\textwidth]{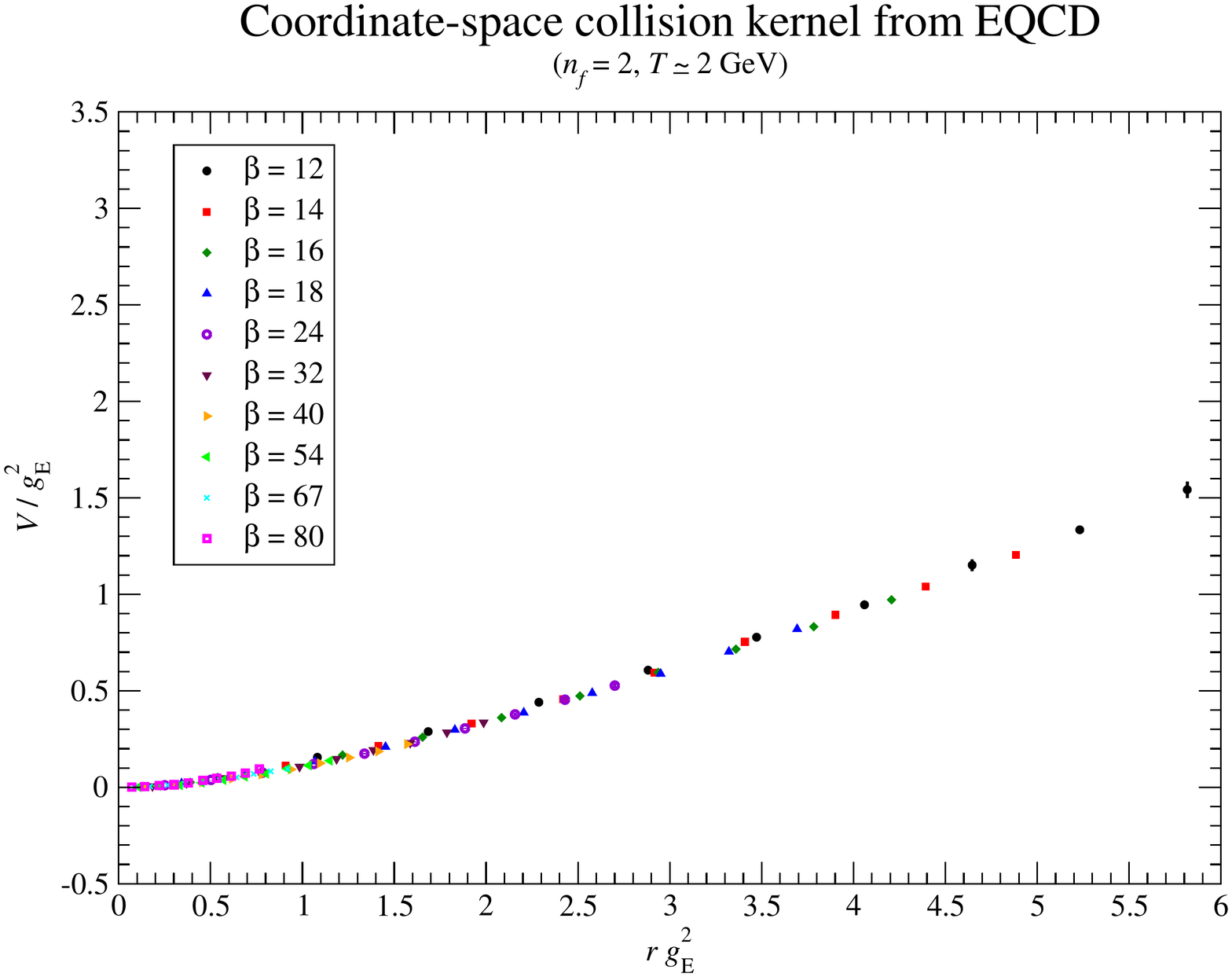}
\caption{Same as in fig.~\protect\ref{fig:colder_decorated_EQCD_V}, but for parameters corresponding to QCD with $n_f=2$ light quark flavors, at a temperature about $2$~GeV.}
\label{fig:hotter_decorated_EQCD_V}
\end{figure*}

At each value of the parameters, we extracted $V(r)$ from the slope of the asymptotic decay of the logarithm of the expectation value of $ W(\ell, r) $ at large $\ell$, according to eq.~(\ref{potential}). This computation was repeated for different values of $n_r$ in independent simulations, to avoid undesired cross-correlations. Fig.~\ref{fig:colder_decorated_EQCD_V} shows the results we obtained from simulations at different $\beta$ values from $12$ to $80$, for the parameter sets listed in the $x_{\mbox{\tiny{L}}}=0.1$ entries of tab.~\ref{tab:parameters}. For these ensembles, the coefficients of the scalar-field terms in eq.~(\ref{lattice_action}) are chosen in such a way, that the theory corresponds to the dimensional reduction of QCD with $n_f=2$ light flavors, at a temperature about $398$~MeV, approximately twice as large as the temperatures at which the deconfinement crossover takes place (for QCD with $n_f=2$ light flavors).\footnote{For details and for a thorough discussion of various technical aspects, including those related to finite-volume effects and other systematics, see ref.~\cite{Hietanen:2008tv} and references therein.} This value is phenomenologically interesting, being in the ballpark of (estimated) temperatures reached at RHIC experiments. However, at this temperature the separation between hard, soft and ultrasoft scales may be only partial, since the coupling is not very small. Hence, we also carried out a set of simulations at a higher temperature, around $2$~GeV (i.e. approximately ten times as large as the deconfinement temperature for $n_f=2$ QCD), corresponding to the $x_{\mbox{\tiny{L}}}=0.06$ data sets in tab.~\ref{tab:parameters}: the results of these simulations are shown in fig.~\ref{fig:hotter_decorated_EQCD_V}. 

In the plots, both $V$ and $r$ are expressed in the appropriate units of $\gsqE$, and we used a tree-level improved distance definition $d_I$ for $r$ (see, e.g., ref.~\cite{Necco:2001xg, Luscher:1995zz} and references therein for a detailed discussion, for the case of Wilson loops in Yang-Mills theory) as well as appropriate renormalization factors~\cite{Moore:1997np}. The very good scaling properties of our data are manifest: results obtained in a rather large range of lattice spacings (spanning almost one order of magnitude), denoted by symbols of different colors, fall on the same curve.

In order to extract the soft contribution to $\qhat$, our lattice results can then be fitted to a functional form that describes the dependence of $V$ on the transverse separation $r$. Perturbation theory~\cite{CaronHuot:2008ni, Ghiglieri:2013gia} suggests that the expansion of $V(r)$ at distances $r$ much shorter than the inverse of the soft scale (but still longer than the inverse of the hard thermal scale $\pi T$) should be of the form
\begin{equation}
\label{perturbative_V}
V/\gsqE = A r \gsqE + B (r \gsqE)^2 + \dots .
\end{equation}

Note that eq.~(\ref{perturbative_V}) includes neither a constant ($V$ has a smooth zero limit for $r \to 0$) nor a logarithmic term: for $p_\perp \gg \mE^2$, the leading-order contribution to $C(p_\perp)$ is $O(\mE^2/p_\perp^4)$, due to mutual cancellation of terms $O(1/p_\perp^2)$. 

Given that the jet quenching parameter is the second moment associated with the $C(p_\perp)$ kernel, and, in turn, the latter is related to $V(r)$ via a Fourier transform, according to  eq.~(\ref{potential_and_C}), $\qhat$ is associated with the curvature of $V(r)$ near the origin, and thus proportional to the parameter $B$ appearing in eq.~(\ref{perturbative_V}). More precisely, $4B=\qhat/\gE^6$. Note, however, that the definition of $B$ is somewhat ambiguous, since, in principle, the right-hand side of eq.~(\ref{perturbative_V}) could also include a term proportional to $(r \gsqE)^2 \ln ( r \gsqE )$ and/or other similar terms.

The NLO weak-coupling computation carried out in ref.~\cite{CaronHuot:2008ni} (see also refs.~\cite{Laine:2012ht, Ghiglieri:2013gia}) predicts $B = \Cfund \Cadj [3 \pi^2 + 10 - 4 \ln (2) ] \mE /(128 \pi^2 \gsqE)$. For the two temperatures investigated in this work, following ref.~\cite{Laine:2009dh} one can estimate the $\mE / \gsqE$ ratio to be around $0.7$ (for our lower-temperature ensemble) or $0.9$ (for the higher-temperature ensemble) in QCD with $n_f=2$ light quark flavors; consistent values are obtained if one uses the physical couplings estimated in ref.~\cite{Laine:2005ai}. This leads to values of $\qhat/\gE^6$ in the ballpark of $0.3$ to $0.4$.

However, fitting our lattice data to the functional form\footnote{In order to compare our EQCD results with those in MQCD discussed in ref.~\cite{Laine:2012ht}, it is particularly convenient to follow an approach as close as possible to the one used in that study, which, in particular, leads to estimate the non-perturbative soft contribution to $\qhat/\gE^6$ in terms of the parameters fitted according to eq.~(\ref{V_fit}), as
\begin{equation}
\label{adapted_equation}
\left\{ 4B + 2C \left[ 2 + \gamma + 4 \ln (r_0 \gsqE / \sqrt{2}) \right] \right\} g^6_{\mbox{\tiny{E}}},
\end{equation}
where $\gamma=0.5772156649\dots$ is the Euler-Mascheroni constant, while $r_0$ denotes Sommer's scale~\cite{Sommer:1993ce} and $r_0 \gsqE \simeq 2.2$~\cite{Luscher:2002qv}.}
\begin{equation}
\label{V_fit}
V/\gsqE = A r \gsqE + B (r \gsqE)^2 + C (r \gsqE)^2 \ln ( r \gsqE ),
\end{equation}
we find numerical evidence for significantly larger values of $\qhat/\gE^6$. For example, for the colder ensemble studied, fitting all our data in the distance interval $0.2 \le r \gsqE \le 0.6$ (in which a finite lower fitting range limit is imposed, in order to mitigate contamination from lattice artifacts) leads to $\qhat/\gE^6 \sim 0.8$---a result dominated by $B \simeq 0.114(7)$, while both $C \simeq 0.05(4)$ and $A \simeq 0.014(17)$ are much smaller and poorly determined. Although this type of analysis is only meant to be suggestive, subtracting the estimated NLO perturbative contribution one finds that the remaining soft contributions to $\qhat$, which are purely non-perturbative, may be in the ballpark of $0.5 \gE^6$, i.e. comparable to the NLO one (which, in turn, at these temperatures appears to be numerically dominant over the LO term~\cite{CaronHuot:2008ni}). 

These values can be converted to physical units: assuming two-loop perturbative results for the coupling~\cite{Laine:2005ai} to be sufficiently accurate even down to the lower of the two temperatures considered (which may not necessarily be the case), one obtains a final estimate for $\qhat$---including also the terms computed perturbatively in ref.~\cite{Arnold:2008vd}---in the ballpark of $6$~GeV$^2$/fm for $T \simeq 398$~MeV. 

\begin{figure*}
\centering
\includegraphics[width=1.\textwidth]{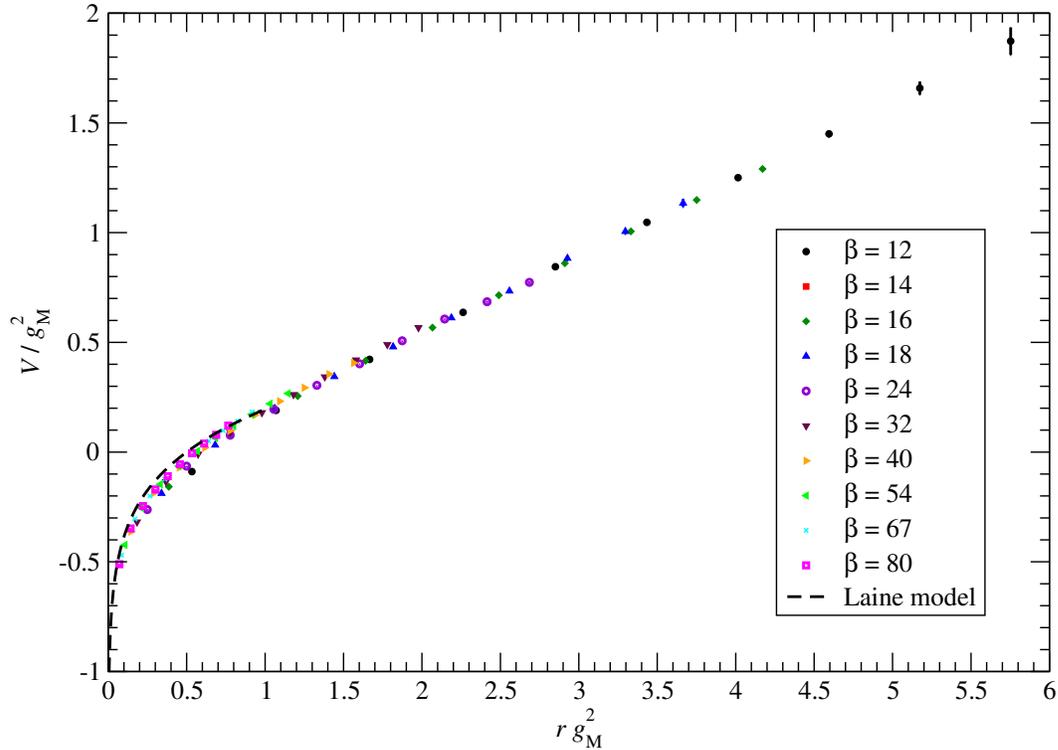}
\caption{As discussed in refs.~\cite{Benzke:2012sz, Laine:2012ht} the static quark-antiquark potential extracted non-perturbatively from simulations in magnetostatic QCD (i.e. in zero-temperature $\SU(3)$ Yang-Mills theory in three Euclidean dimensions) encodes information on the contribution to the jet quenching parameter from the ultrasoft scale. This plot shows our results from a new set of simulations, in comparison with a curve corresponding to the modelling of data from ref.~\cite{Luscher:2002qv} performed by Laine in ref.~\cite{Laine:2012ht} (dashed line). The data sets---as well as the continuum model described by the curve---are required to go through the point of coordinates $ 2 r g^2_{\mbox{\tiny{M}}} = 1$, $V=0$---except those on the coarsest lattices, which are matched in the linear-rise region.}
\label{fig:ordinary_MQCD_V}
\end{figure*}

Interestingly, the rather large non-perturbative effects we observed seem to be genuinely due to soft physics, given that the analysis of analogous contributions from the ultrasoft sector carried out in ref.~\cite{Laine:2012ht} found much smaller results. In order to understand if this difference may be (at least partially) due to discretization effects, we also repeated the analysis carried out in ref.~\cite{Laine:2012ht} in the MQCD setup, extending it to data obtained from lattices much finer than those considered in ref.~\cite{Luscher:2002qv}. To this purpose, we performed a set of high-precision measurements of Wilson loops in zero-temperature $\SU(3)$~Yang-Mills theory in three Euclidean dimensions, for the same values of the Wilson lattice parameter $\beta$ that we used in our EQCD simulations,\footnote{As we mentioned in sect.~\ref{sec:setup}, the gauge couplings in EQCD and in MQCD coincide, up to subleading corrections: $\gsqE \simeq \gsqM$.} i.e. from $12$ to $80$, and for the same volumes and comparable statistics.\footnote{Note that, by construction, MQCD does not have parameters which correspond to different temperatures (or number of quark flavors) of the original, four-dimensional theory.} The parameters of the corresponding data sets are given in the $x_{\mbox{\tiny{L}}}=0$ entries of tab.~\ref{tab:parameters}. As compared to ref.~\cite{Luscher:2002qv}, this much broader range of gauge couplings results in a fourfold increase in the lattice cutoff, enabling us to have much better control of the continuum limit, and very precise and accurate data in the range of distances of interest for the present analysis. Our results from this study are shown in fig.~\ref{fig:ordinary_MQCD_V}, where we also display a curve, which would correspond to a potential parameterized according to the fit carried out in ref.~\cite{Laine:2012ht}.\footnote{The data and the curve are constrained to cross the horizontal axis at $2r g^2_{\mbox{\tiny{M}}} = 1$ (for the data sets, a linear interpolation of the two nearest points is used, at each $\beta$~value)---except for the coarsest lattices, for which the matching is done in the region where $V$ is linearly rising with $r$.} The figure clearly shows that the curve is compatible with the continuum limit extrapolated from our data. In particular, our results confirm that the data modelling reported in ref.~\cite{Laine:2012ht} is also valid down to distances shorter than those investigated in ref.~\cite{Luscher:2002qv} (on which the analysis of ref.~\cite{Laine:2012ht} is based). Note that, in principle, with our data one could also try and test the perturbative behavior predicted for MQCD in ref.~\cite{Pineda:2010mb}. However, since lattice results at very short distances, corresponding to just a few lattice spacings, may be affected by non-negligible discretization effects, in the present work we did not perform this test, and restricted our whole analysis to data at distances $r \gsqM \ge 0.2$.

Coming back to our EQCD results, since the numerical values obtained from the crude analysis that we carried out above are somewhat dependent on the fitting procedure details, do not involve a proper continuum extrapolation of the lattice results, nor an estimate of the systematic uncertainties due to the lack thereof, we performed a more direct comparison of our data with the perturbative NLO predictions for $V$, in order to get a better physical interpretation for the origin of this discrepancy. To this purpose, for each data set corresponding to a fixed temperature ensemble and a given lattice spacing, we performed an interpolation of our values for $V$ as a function of $r$ using cubic splines and attempted an estimate of the continuum limit curve (and associated errorband) at each of the interpolated $V(r)$ values.\footnote{Since, for each of the two temperature ensembles and for each value of $r$, the interpolated data points obtained from simulations at different lattice spacing are essentially mutually compatible (within uncertainties estimated with the jackknife method), at each $r$ value we defined the errorband of the continuum curve as the envelope of the data and errorbars at that point, and the continuum values as the central values of the resulting errorband. This procedure turns out to be more reliable than a direct, pointwise $a \to 0$ extrapolation (especially at distances where only a limited number of data points are available).} 

The continuum curve (together with the lattice data from which it was obtained) was then plotted in units of the ``natural'' soft scale, i.e. the Debye mass $\mD$, and compared with the corresponding NLO curve in continuum, obtained from the collision kernel $C(p_\perp)$~\cite{CaronHuot:2008ni} via a numerical Fourier transform, see eq.~(C.4) in ref.~\cite{Ghiglieri:2013gia}.\footnote{We thank Jacopo~Ghiglieri and Aleksi~Kurkela for providing us with the numerical values for that expression.}

As mentioned above, at this order the perturbative expression for the Debye mass coincides with $\mE$:
\begin{equation}
\label{perturbative_Debye_mass}
\mD = gT \sqrt{\frac{N + n_f \tfund }{3}} + O(g^2 T).
\end{equation}
Unsurprisingly, plotting $V$ and $r$ as a function of $\mD$ according to the definition in eq.~(\ref{perturbative_Debye_mass}), we found that the lattice data deviate strongly from the perturbative curve, and that, in particular, they exhibit much more pronounced curvature near the origin.

However, we also discovered that, at both temperatures, the agreement with the NLO perturbative curves improves dramatically (becoming, in fact, nearly perfect), provided one uses non-perturbatively defined values for $\mD$. In fact, it is well known that, for all temperatures of phenomenological relevance at collider experiments, the Debye mass receives large contributions from terms which are neglected in eq.~(\ref{perturbative_Debye_mass}). Following the approach proposed in ref.~\cite{Arnold:1995bh}, the non-perturbative $O(g^2 T)$ contribution\footnote{Further $O(g^3 T)$ contributions are suppressed by very small coefficients~\cite{Kajantie:1997pd}.} to $\mD$ was computed on the lattice in ref.~\cite{Laine:1999hh}, where it was found that
\begin{equation}
\label{nonperturbative_Debye_mass}
\mD \simeq gT \sqrt{\frac{N + n_f \tfund }{3}} + 1.65 g^2 T ,
\end{equation}
so that, at the temperatures considered here, the numerical value of $\mD$ is actually dominated by the $O(g^2 T)$ term (in spite of the fact that the latter is parametrically suppressed by one power of $g$).

\begin{figure*}
\centering
\includegraphics[width=1.\textwidth]{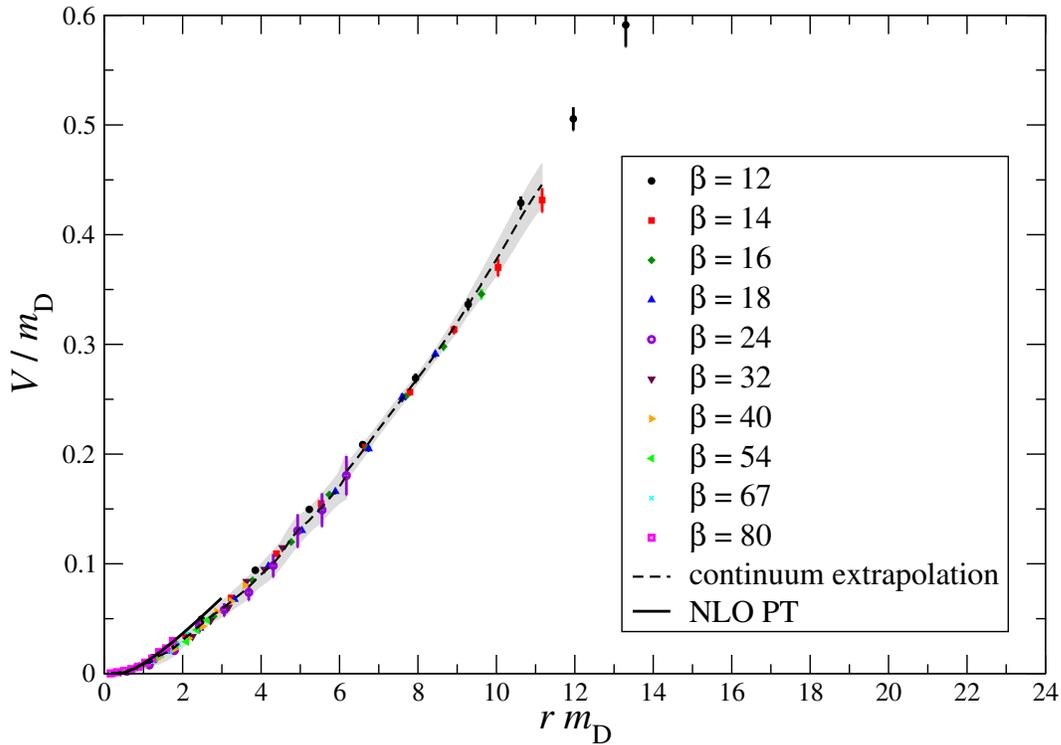}
\caption{Comparison of our lattice results for the EQCD collision kernel (whose continuum extrapolation is shown by the dashed line, with the gray band denoting the corresponding uncertainty) with the next-to-leading order perturbative prediction~\cite{CaronHuot:2008ni, Ghiglieri:2013gia} (solid black line), using the non-perturbative values of the Debye mass $\mD$ obtained in ref.~\cite{Laine:1999hh}. This plot shows the results for our ensemble at the lower temperature, $T \simeq 398$~MeV, corresponding to $g^2 \simeq 2.6$~\cite{Laine:2005ai}.}
\label{fig:mD_colder_decorated_EQCD_V}
\end{figure*}

\begin{figure*}
\includegraphics[width=1.\textwidth]{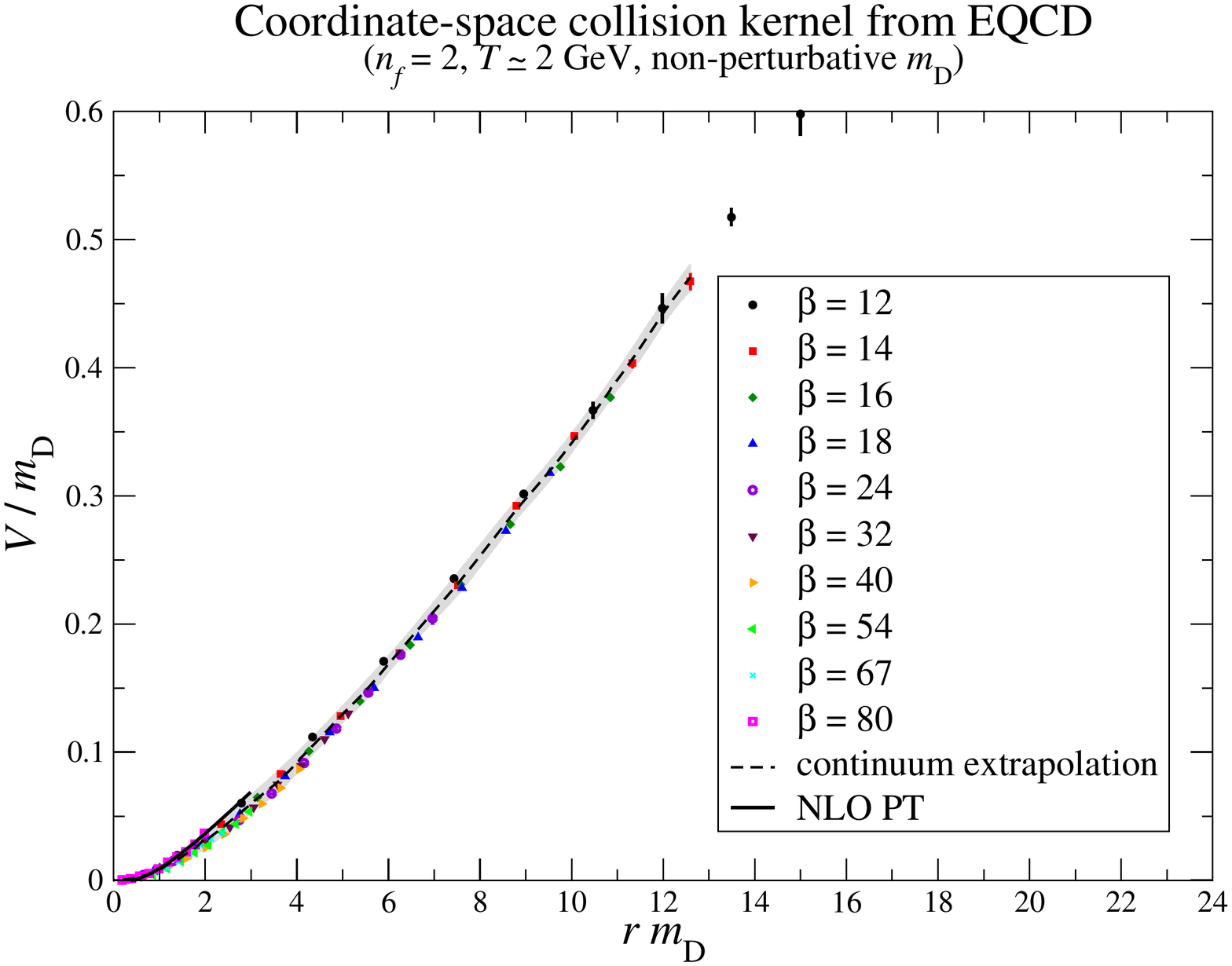}
\caption{Same as in fig.~\protect\ref{fig:mD_colder_decorated_EQCD_V}, but for our higher-temperature ensemble, at $T \simeq 2$~GeV, corresponding to $g^2 \simeq 1.5$~\cite{Laine:2005ai}.}
\label{fig:mD_hotter_decorated_EQCD_V}
\end{figure*}

Expressing our results for $V/\mD$ \emph{versus} $r \mD$ using the non-perturbatively evaluated Debye mass given in eq.~(\ref{nonperturbative_Debye_mass}), we observe that our lattice results come to agree very well with the corresponding NLO perturbative expression, as shown in fig.~\ref{fig:mD_colder_decorated_EQCD_V} (for our colder ensemble) and in fig.~\ref{fig:mD_hotter_decorated_EQCD_V} (for the data set at higher temperature). In these two figures we also display the curves obtained by extrapolation of our lattice results to the continuum limit, which are given by the dashed lines (with uncertainties denoted by the gray bands). 

As fig.~\ref{fig:mD_colder_decorated_EQCD_V} and fig.~\ref{fig:mD_hotter_decorated_EQCD_V} show, using the non-perturbative values of $\mD$ from ref.~\cite{Laine:1999hh} at physical couplings~\cite{Laine:2005ai}, our data are in excellent agreement with the perturbative next-to-leading order curve~\cite{CaronHuot:2008ni, Ghiglieri:2013gia} in the region of interest,  $r \mD \lesssim 1$. Note that at very large distances the scalar field effectively decouples, so that the dynamics of ultrasoft modes can be described by pure Yang-Mills theory in three spatial dimensions (i.e. MQCD). In that limit, the $W$ operator goes over to the trace of a purely spatial Wilson loop, and $V$ approaches the potential of 3D Yang-Mills theory. In the opposite limit, the NLO PT calculation reported in refs.~\cite{CaronHuot:2008ni, Ghiglieri:2013gia} predicts that $V(r)$ starts from zero at $r=0$, and that its small-$r$ behavior includes a linear term with \emph{negative} slope. With our simulations, we are unable to reliably probe those very short distances, due to the finiteness of our lattice spacings.\footnote{As we mentioned above, for the colder ensemble our fit result for the coefficient of the linear term in eq.~(\ref{V_fit}) is positive, but smaller than its uncertainty, $A=0.014(17)$. The slight tension with the perturbative continuum prediction $-\Cfund \Cadj / (32 \pi)$ (at the level of two to three standard deviations) can probably be explained in terms of systematic uncertainties related to the fitting procedure.} In any case, at very short distances the theory ceases to be a valid effective description for finite-temperature QCD, since, by construction, it neglects the hard thermal modes, with momenta $O(\pi T)$, of the full theory. In fact, as discussed in ref.~\cite{Ghiglieri:2013gia}, one can show that the negative linear term appearing in the NLO PT expression of $V(r)$ in EQCD exactly cancels against an analogous contribution with the opposite sign, which appears in the soft limit of the hard contribution to $\qhat$~\cite{CaronHuot:2008ni}. Finally, fig.~\ref{fig:mD_colder_decorated_EQCD_V} and fig.~\ref{fig:mD_hotter_decorated_EQCD_V} also reveal that the data sets at the two different temperatures fall on the same curve (within uncertainties) once they are expressed in terms of the non-perturbatively evaluated Debye mass $\mD$: this indicates that, essentially, the temperature dependence of $V$ is inherited from $\mD$.

The fact that at short to intermediate distances the continuum curve obtained from our lattice data agrees almost perfectly with the NLO PT curve, \emph{provided the non-perturbative value of the Debye mass is used}, suggests an alternative way to estimate the soft contribution to $\qhat$, by replacing the non-perturbative $\mD$ value in the analytical NLO prediction for the jet quenching parameter. This results in
\begin{equation}
\label{qhatNLO}
\qhatsoft = g^4 T^2 \mD \Cfund \Cadj \frac{ 3 \pi^2 + 10 - 4 \ln 2 }{32 \pi^2}.
\end{equation}
With the coupling value used above ($g^2 \simeq 2.6$ at $T \simeq 398$~MeV), eq.~(\ref{qhatNLO}) leads again to a final value for $\qhat$ in the ballpark of $6$~GeV$^2$/fm for RHIC, consistently with our determination with the previous method. Note that at this temperature the NLO result is much larger than the leading-order one (which, at these temperatures, can be estimated to be of the order of $1$~GeV$^2$/fm~\cite{Arnold:2008vd}): see ref.~\cite{CaronHuot:2008ni} for a discussion.

The error budget in our computations includes uncertainties of statistical and of systematic nature. As the plots show, statistical errors are rather small. Our approach allows us to study the soft physics of QCD with light dynamical quarks by simulations of a purely bosonic effective theory, and our implementation of the multilevel algorithm~\cite{Luscher:2001up} for the operator under consideration tames the problem of exponential decay in the signal-to-noise ratio for this non-local operator. In view of the high statistical precision of our data, the error budget of our results is likely dominated by systematic errors, which have to be estimated accurately.

Potential sources of systematic uncertainties affecting our calculation include:
\begin{itemize}
\item effects due to incomplete separation between the hard and soft scales,
\item finite-volume corrections,
\item artifacts due to the finiteness of the lattice cutoff, and
\item uncertainties related to the fitting procedure of our data to eq.~(\ref{perturbative_V}).
\end{itemize}

The effects related to incomplete separation between hard and soft scales in the plasma are dependent on the value of the coupling $g$, hence on the temperature, of the four-dimensional theory. Hard, soft and ultrasoft scales differ parametrically by powers of $g$, hence they become well separated only in the weak-coupling limit. Due to the logarithmic running of the coupling, this means that a clear-cut separation of such scales could only be achieved at very high temperatures, out of the reach of present accelerator experiments. However, by now there exists a large body of literature (see refs.~\cite{Kajantie:2002wa, Laine:2005ai, Hietanen:2008tv} and references therein) showing that analytical computations relying on this separation of scales are actually quite accurate even at surprisingly low temperatures---possibly down to twice the deconfinement temperature~\cite{Borsanyi:2012ve}. While quantifying the systematic uncertainties due to partial scale overlap in our present work is nearly impossible (because they are generally observable-dependent, and we have no alternative, independent non-perturbative first-principle computation to compare our results with), numerical evidence from the aforementioned studies led us to choose twice the deconfinement temperature as the physical temperature of our ``colder'' ensemble. Systematic effects due to incomplete scale separation are, of course, expected to be milder for our ``hotter'' ensemble: its physical temperature (approximately five times larger than that of the colder ensemble) is well inside the regime where, for example, lattice studies of the equation of state~\cite{Borsanyi:2012ve, Boyd:1996bx} are in very good agreement with perturbative computations relying on the separation between the hard, soft and ultrasoft scales~\cite{Kajantie:2002wa}. As a final remark on this issue, it is perhaps worth noting that the actual ratio of the soft to hard---as well as ultrasoft to soft---scales is $g/\pi$ (rather than $g$), indicating that the minimal temperatures at which scale separation still works may be lower than na\"{i}vely expected.

General aspects of finite-volume effects in lattice simulations of EQCD have already been discussed in detail in the literature (see ref.~\cite{Hietanen:2008tv} and references therein), hence we will not repeat the whole discussion here. Suffice it to say, that the 3D $\SU(3)$ gauge theory coupled to a scalar field in the adjoint representation is confining and the lightest physical state in the spectrum has a finite mass~\cite{Hart:2000ha}, so that finite-volume effects are exponentially suppressed with the linear size of the system. In particular, previous numerical studies, carried out in ref.~\cite{Hietanen:2008tv} at the same parameters used in our simulations, showed explicitly that no finite-volume effects are visible for lattices of linear extent (in lattice spacing units) larger than $\beta$. As tab.~\ref{tab:parameters} shows, this requirement is lavishly satisfied in our simulations. In addition, all Wilson loops studied are much smaller than the lattice cross-section, and we observed no evidence whatsoever of finite-volume effects in our data. This leads us to conclude that finite-volume effects are negligible in the total error budget of our computation.

As we mentioned above, in order to reduce contamination by finite-cutoff artifacts, in our study we did not include data corresponding to distances of only one or a few lattice spacings, and used appropriate lattice definitions of the distances, together with known renormalization factors. In principle, the lattice discretization of the $W$ operator considered in this work could pose additional subtleties (related, in particular, to the presence of parallel transporters along the real-time direction). The numerical evidence from our results, however, suggests that no major renormalization effects are present. In particular, the perturbative analysis in the continuum~\cite{CaronHuot:2008ni} shows that the short-distance behavior of $V$ is determined by a partial \emph{cancellation} between gauge and scalar field propagators at momenta much larger than $\mD$, and the fact that our lattice results for $V$ tend to zero for $r \to 0$ (and closely follow the analytical prediction, once the leading non-perturbative correction to the Debye mass is taken into account) gives confidence that our lattice implementation of the $W$ operator is not affected by strong artifacts. For an analytical study of the renormalization properties of this operator, see ref.~\cite{D'Onofrio:2014qxa}.

Systematic uncertainties related to the fitting procedure in our estimate of $\qhat$ from fits to eq.~(\ref{V_fit}) appear to be non-negligible, since the fitted parameters exhibit some dependence on the fitting range, and (to a lesser extent) on the inclusion of possible subleading corrections. However, the final results for $\qhat$ are relatively stable (this is probably due to the fact that the terms related to the curvature of $V$ in our data are not strongly dependent on the fitting range, even for $r \gsqE \gtrsim 1$).

In view of this error budget, we believe that a rough estimate of the total relative uncertainty affecting our final results for $\qhat$ is around $15$ to $20 \%$.

It is intriguing to note that the value for $\qhat$ that we obtained at the lower temperature simulated (comparable to those realized at RHIC) is rather close to the estimate from a holographic computation carried out in ref.~\cite{Liu:2006ug} for the $\mathcal{N}=4$ supersymmetric plasma.\footnote{As it is well known, many of the \emph{qualitative} differences between the $\mathcal{N}=4$ supersymmetric Yang-Mills theory and QCD at zero temperature disappear (or are at least mitigated) in the deconfined phase at high temperature. Hence it may make sense to compare holographic predictions for the $\mathcal{N}=4$ supersymmetric plasma with QCD: examples of such comparisons include those discussed in refs.~\cite{Gubser:1996de, Gubser:1998nz, Policastro:2001yc, Kovtun:2004de, Herzog:2006gh, Gubser:2006bz, Gubser:2006qh} (see also refs.~\cite{Son:2007vk, Mateos:2007ay, Shuryak:2008eq, Gubser:2009md, CasalderreySolana:2011us} for a discussion).} That computation was done in the 't~Hooft and strong coupling limits, and numerical estimates were obtained by plugging the number of colors $N=3$ and a value of the 't~Hooft coupling $\lambda \simeq 6 \pi$ in the final formula
\begin{equation}
\label{qhat_SYM}
\qhat = \frac{\Gamma(3/4)}{\Gamma(5/4)}\sqrt{\pi^3\lambda}T^3  .
\end{equation}
This led to values for $\qhat$ between $4.5$ and $10.6$~GeV$^2$/fm for temperatures in the range between $300$ and $400$~MeV. In fact, using a value of the coupling closer to the one used for our conversion to physical units at temperatures close to $400$~MeV, the prediction in eq.~(\ref{qhat_SYM}) goes down to approximately $6.7$~GeV$^2$/fm, in even better agreement with our result. 

Very close results were also obtained in ref.~\cite{Armesto:2006zv}, where it was shown that the inclusion of the first finite-$\lambda$ correction would slightly reduce the numerical value for $\qhat$. For values of the coupling relevant for comparison with our data,  the reduction is of the order of a few percent. On the other hand, lower $\qhat$ values were obtained in ref.~\cite{Gursoy:2009kk}, which discusses computations in the improved holographic QCD model proposed in refs.~\cite{Gursoy:2007cb, Gursoy:2007er}.

A comparison with experimental RHIC results (or with phenomenological models involving some form of experimental input) also involves a number of subtleties: in particular, the estimates quoted in the literature are to be interpreted as values averaged over time, as the plasma expands and cools down. Even with these caveats, it is interesting to note that our result compares favorably with estimates from some phenomenological computations using input from RHIC experiments, which predict values around $5$ to $15$~GeV$^2$/fm~\cite{Dainese:2004te, Eskola:2004cr}. On the other hand, the comparison of different formalisms carried out in the more recent ref.~\cite{Bass:2008rv} seems to point towards smaller values (albeit with somewhat large uncertainties, and with some dependence on the choice of the underlying phenomenological description).

As for our results at the higher temperature (about $2$~GeV, i.e. approximately one order of magnitude larger than the temperature at which deconfinement takes place), where perturbation theory should work more accurately, using, again, coupling values from  ref.~\cite{Laine:2005ai}, we obtain that the contribution to $\qhat$ from soft physics is around $2 T^3$. While at this higher temperature it is expected that hard, soft and ultrasoft scales are more clearly separated, a comparison with holographic computations is less motivated, since the latter are based on the supergravity approximation, which relies on strong-coupling assumptions. On the other hand, it would be interesting to see if this prediction could be confirmed in future experiments.

\section{Discussion and conclusions}
\label{sec:discussion}

In this work, we carried out a first-principle theoretical computation of the soft-scale contribution to the momentum broadening of an ultrarelativistic parton (specifically: a light quark) moving through the deconfined state of strongly interacting matter at high temperature.

We emphasize that our approach to this problem is systematic, and that it is not model-dependent. Although the investigation of phenomena involving real-time dynamics by simulations on a Euclidean lattice is a notoriously difficult problem, for the case considered in this work it was possible, thanks (in particular) to the peculiar properties of plasma fields at the soft scale $gT$. As initially suggested in ref.~\cite{CaronHuot:2008ni}, the contribution to hard-parton momentum broadening from collisions with plasma constituents involving momentum transfers of order $gT$ is essentially insensitive to the precise value of the parton velocity---and such would remain, even in the unphysical limit of a superluminal parton. This implies that the corresponding contribution can be computed in the Euclidean setup of a dimensionally reduced effective theory, capturing the physics of long-wavelength plasma modes, i.e. in electrostatic QCD. This observation opened up the possibility of addressing the study of a certain class of dynamical, real-time phenomena by means of standard computations on a Euclidean lattice.

Following the formalism reviewed in ref.~\cite{CasalderreySolana:2007zz}, the phenomenon studied in this work can be described in terms of the two-point correlation function of spatially separated Wilson lines on the light cone. After briefly reviewing the rigorous demonstration~\cite{CaronHuot:2008ni, Ghiglieri:2013gia} that the study of this correlation function can be reduced to a Euclidean setup, we presented a numerical study of the soft contribution to the jet quenching parameter $\qhat$ in the Wilson lattice regularization of EQCD.\footnote{One subtle but important aspect deserves emphasis~\cite{CaronHuot:2008ni}: physically, the mathematical procedure by which the correlator of light-cone Wilson lines can be mapped to a Euclidean correlator corresponds to identifying the limits in which the parton velocity tends to the speed of light from below or from above. The equivalence of these two limits is guaranteed for classical plasma physics effects, which is the case relevant for the problem under consideration.}

Our computation is based on the non-perturbative numerical evaluation of vacuum expectation values of a gauge-invariant operator in EQCD, which represents the dimensionally reduced counterpart of a Wilson loop with two spatially separated sides lying along a light-cone direction in the full theory in $(3+1)$-dimensional Minkowski spacetime.\footnote{For a recent study of this operator in a classical lattice gauge theory formulation, see refs.~\cite{Laine:2013lia, Laine:2013apa}.} In the Wilson lattice regularization, it can be expressed in terms of a ``decorated'' Wilson loop, which includes insertions of \emph{Hermitian} matrices (representing parallel transporters along the \emph{real-time} direction, and obtained by exponentiation of the adjoint scalar field of EQCD). Note that this approach allows one to directly access the Fourier transform of the collision kernel $C(p_\perp)$ appearing in eq.~(\ref{qhat_definition}), i.e. the differential collisional rate describing interactions between the plasma and the hard parton. This quantity is ``more fundamental'' than the jet quenching parameter itself---which can be derived from it, but whose definition is affected by ambiguities, including, in particular, those related to the definition of the upper bound on the $p_\perp$ modulus in the integration in eq.~(\ref{qhat_definition}).

Our construction closely follows the corresponding formulation in the continuum~\cite{CaronHuot:2008ni}. Note that the latter implies, in particular, that at LO at weak coupling, a cancellation between gauge and scalar field propagators takes place at large $p_\perp$ (see also ref.~\cite{Aurenche:2002pd}). In turn, this leads to the absence of constant and Coulomb terms in the expression for the coordinate-space transverse collisional kernel. Our numerical results at short distances confirm these expectations. 

One way to estimate the contribution to $\qhat$ at the soft scale consists, then, in following the procedure used in ref.~\cite{Laine:2012ht} for the analysis of the contribution from the ultrasoft scale, using the fact that $\qhat$ is related to the curvature of $V$ near the origin. 

In the present work, we studied jet quenching in QCD with two light quark flavors, focusing on two different temperatures, respectively close to $0.4$ and $2$~GeV. The higher of these two temperatures is already (well) inside the regime where, for example, computations of the equation of state exploiting the separation of scales between hard, soft and ultrasoft physics do work well~\cite{Kajantie:2002wa, Laine:2005ai}, and is of the same order of magnitude as the temperatures that may be reached in future collider experiments.\footnote{For example, a heavy-ion collision program at the (hypothetical) Very Large Hadron Collider could probably probe temperatures around or beyond $1$~GeV.} The lower temperature, on the other hand, was chosen because it is in the regime probed experimentally at RHIC and LHC, and, hence, it is of more immediate phenomenological interest. The separation between hard, soft and ultrasoft scales at such a ``low'' temperature (approximately twice as large as the temperature at which the theory deconfines) is not obvious, but previous works give some indication that probably this temperature is still within (or at the boundary of) the region where effective field theories relying on such separation of scales do work. As an example, a recent, high-precision non-perturbative lattice study of the equation of state in $\SU(3)$ Yang-Mills theory over a very large temperature range, reported in ref.~\cite{Borsanyi:2012ve}, showed good agreement with analytical computations relying on such separation of scales, all the way down to temperatures close to $2 T_c$. In addition, one further motivation for our choice of the lower temperature consists in the fact that, as the coupling is relatively strong, it is interesting to compare our results with predictions from holographic computations~\cite{Liu:2006ug, Armesto:2006zv, Gursoy:2009kk}.

Our results indicate that non-perturbative soft contributions to $\qhat$ are large, and likely play a dominant r\^ole in the momentum broadening of a hard parton in the QGP. At the lower temperature that we considered, around $400$~GeV, our final estimate for $\qhat$, including the perturbatively known contributions, is about $6$~GeV$^2$/fm, with an estimated uncertainty of the order of $15$ to $20 \%$. This value is comparable to estimates based on holographic methods~\cite{Liu:2006ug, Armesto:2006zv, Gursoy:2009kk} as well as on certain model computations with phenomenological input~\cite{Dainese:2004te, Eskola:2004cr}---although more recent works in this direction favor somewhat lower values~\cite{Bass:2008rv}.

We can also compare our results with related works in the recent literature~\cite{Benzke:2012sz, Laine:2012ht, Laine:2013lia, Laine:2013apa}. As discussed above, the most closely related work is reported in ref.~\cite{Laine:2012ht}, which extracted the non-perturbative contribution to $\qhat$ from the ultrasoft sector, using lattice results from ref.~\cite{Luscher:2002qv}. As we mentioned in sect.~\ref{sec:lattice}, we also carried out a new set of MQCD simulations, extending the results reported in ref.~\cite{Luscher:2002qv} to smaller distances and finer lattices. These results confirm the validity of the fit performed in ref.~\cite{Laine:2012ht} and the conclusions reached in that work, that contributions to $\qhat$ from the purely chromomagnetic sector are numerically subleading with respect to the perturbative ones, and that a non-perturbative determination including effects from the chromoelectric sector is well motivated. Such determination was carried out in the present work.

Another recent work using data from ref.~\cite{Luscher:2002qv} to estimate ultrasoft contributions to $\qhat$ was reported in ref.~\cite{Benzke:2012sz}, in which the potential extracted non-perturbatively in $\SU(3)$~Yang-Mills theory in three dimensions was parameterized in a different way---namely, in a series in inverse powers of the distance. While this parametrization is known to provide a good modelling of the confining potential at intermediate to long distances (see, e.g., section~2 in ref.~\cite{Panero:2005iu} and references therein for a discussion), one may argue that perhaps it is not ideally-suited for the problem under consideration, and it implies a stronger dependence of $\qhat$ on the chosen maximum-momentum scale.

Another work closely related to our discussion is ref.~\cite{Laine:2013lia} (see also ref.~\cite{Laine:2013apa}), which presents a study of light-cone Wilson lines in classical lattice gauge theory. Although classical lattice gauge theory is not expected to be quantitatively accurate enough for physics at the soft scale, due to a sensitiveness to discretization effects~\cite{Bodeker:1995pp, Arnold:1997yb}, the authors of ref.~\cite{Laine:2013lia} found convincing evidence supporting the validity of the analytical continuation of the Wilson loop across the light cone (in agreement with the theoretical proof reviewed above), and clear indications that, even at rather short distances, the effect of interactions may be quantitatively more relevant than expected from weak-coupling arguments.

More recently, ref.~\cite{Nam:2014sva} proposed to evaluate $\qhat$ combining information about the temperature dependence of the instanton distribution and non-perturbative lattice results for the confining potential. That computation, however, does not involve real-time Wilson lines.

Finally, a different type of approach to study the jet quenching parameter on the lattice was suggested in ref.~\cite{Majumder:2012sh}. The general strategy discussed in that work is, however, quite different with respect to our present approach, and the preliminary numerical study presented in ref.~\cite{Majumder:2012sh} is carried out in $\SU(2)$ pure Yang-Mills theory, so that a direct comparison with our results is not possible.

To summarize, our lattice study indicates that non-perturbative terms from physics at the soft scale give a rather large contribution to the jet quenching parameter. Although jet quenching belongs to the class of real-time phenomena in thermal QCD, which are notoriously challenging to study on a Euclidean lattice, in this work we could bypass this intrinsic difficulty, by mapping the physics of interest to a dimensionally reduced Euclidean effective theory for hot QCD. The rigorous proof that this is possible is presented in refs.~\cite{CaronHuot:2008ni, Ghiglieri:2013gia}; recently, closely related ideas have also been discussed in ref.~\cite{Brandt:2014uda}, elucidating the relation between screening masses and light-cone rates in real time. Our results indicate that soft contributions to $\qhat$ beyond NLO perturbative terms are around $0.5 \gE^6$ at temperatures currently accessible in accelerator experiments. In turn, including also the other perturbatively known terms, this leads to a final result for $\qhat$ around $6$~GeV$^2$/fm for RHIC temperatures. This value is affected by some uncertainty, which we estimated to be of the order of $15$ to $20 \%$.

Due to the nature of $\qhat$ as a collective transport parameter, when considering phenomenological applications e.g. in hydrodynamics simulations~\cite{Hirano:2003hq, Chaudhuri:2005vc}, it may be appropriate to define the numerical value of $\qmax$, the maximum momentum modulus in the integral defining $\qhat$, according to some criterion related to the specific model or to the specific type of process considered.\footnote{In particular, $\qmax$ may depend on the jet lifetime.} Presently, however, such dependence is not relevant for our discussion, because it leads to variations on our final estimates within the uncertainties on our results.

As for possible generalizations of the present work, it may be interesting to repeat the present computations with an improved formulation of the lattice action, as recently suggested in ref.~\cite{Mykkanen:2012dv}. Another possible interesting generalization would be to study the dependence of the jet quenching parameter on the number of colors $N$. In fact, it is known that the dynamics of gauge theories simplifies considerably at large $N$~\cite{'tHooft:1973jz} (see also refs.~\cite{Lucini:2012gg, Vicari:2008jw, Panero:2012qx, Lucini:2013qja} for recent reviews). In addition, the large-$N$ limit is also important for computations based on the holographic correspondence, including those addressing the jet quenching phenomenon~\cite{Liu:2006ug, Armesto:2006zv, Gursoy:2009kk} with which we compared our results. By now it has been established that several observables relevant for the QCD plasma in equilibrium have only a mild dependence on the number of colors: lattice studies show that the deconfinement temperature~\cite{Lucini:2002ku, Lucini:2003zr, Lucini:2005vg, Datta:2009jn, Lucini:2012wq}, the equation of state~\cite{Panero:2009tv, Bringoltz:2005rr, Panero:2008mg, Panero:2009wr, Datta:2010sq} and Polyakov loops in different representations~\cite{Mykkanen:2012ri, Mykkanen:2011kz} are only weakly dependent on $N$ (up to trivial factors).\footnote{Analogous results have been obtained in $(2+1)$ spacetime dimensions~\cite{Liddle:2008kk, Bialas:2008rk, Caselle:2011fy, Caselle:2011mn, Bialas:2012qz}.} To test whether this also holds for quantities involving genuine real-time dynamics, one could repeat the present computation for $N \ge 3$ colors. In particular, it would be very interesting to investigate the non-trivial dependence of $\qhat$ on the number of colors $N$ found in ref.~\cite{Liu:2006ug} for the strongly coupled $\mathcal{N}=4$ plasma: as eq.~(\ref{qhat_SYM}) shows, the holographic computation predicts $\qhat$ to be independent of $N$ at fixed 't~Hooft coupling. If this is also the case for the QCD plasma at strong coupling, then it follows that the jet quenching parameter cannot be interpreted as a quantity ``measuring'' either the entropy density $s$, or a sort of ``gluon number density'' proportional to $\epsilon^{3/4}$ (where $\epsilon$ denotes the energy density), because in the deconfined phase both $s$ and $\epsilon$ have been shown to scale very precisely with the number of gluon degrees of freedom, i.e. proportionally to $(N^2-1)$, for all temperatures~\cite{Panero:2009tv}. Extending the present work to $N > 3$ could be easily done, following the proper definition of the corresponding dimensionally reduced EQCD and MQCD effective theories~\cite{Kajantie:1997tt, Laine:1997dy}. Finally, a more straightforward generalization of the work presented here would be to repeat the computations at different values of the temperature, in order to investigate the precise dependence of $\qhat$ on $T$ (beyond the expectation $\qhat \propto T^3$, which is based on purely dimensional grounds). We leave these research directions for the future.

\vskip1.0cm 
\noindent{\bf Acknowledgements.}\\
This work is supported by the Academy of Finland, project 1134018, by the Spanish MINECO's ``Centro de Excelencia Severo Ochoa'' programme under grant SEV-2012-0249, by the German DFG (SFB/TR 55), and partly by the European Community under the FP7 programme HadronPhysics3. Part of this work was carried out during the ``Heavy quarks and quarkonia in thermal QCD'' at ECT$^\star$ in Trento, Italy. Part of the numerical simulations was carried out at the Finnish IT Center for Science (CSC) in Espoo, Finland. We thank Francesco~Bigazzi, Dietrich~B\"odeker, Simon~Caron-Huot, Jacopo~Ghiglieri, Aleksi~Kurkela, Mikko~Laine, Berndt~M\"uller, Anne~Mykk\"anen, Carlos~Salgado, Antonio~Vairo and Xin-Nian~Wang for helpful comments and discussions.

\bibliography{jetquenching}

\end{document}